\documentclass[
prc,
aps,
superscriptaddress,
twocolumn,
secnumarabic,
balancelastpage,
longbibliography,
amsmath,
amssymb,
nofootinbib,
floatfix,
reprint
]{revtex4-1}

\usepackage{graphicx}      % tools for importing graphics
\usepackage{bm}            % special bold-math package. usge: \bm{mathsymbol}
\usepackage{amsmath}
\usepackage{amssymb}
\usepackage{lipsum}
\usepackage[version=3]{mhchem}
\usepackage[colorlinks=true]{hyperref} 
\usepackage{siunitx}

\newcommand{\mmicro}{\si\micro}  
\DeclareMathOperator\erfc{erfc}

\newcommand\Tstrut{\rule{0pt}{2.6ex}}         % = `top' strut
\newcommand\Bstrut{\rule[-0.9ex]{0pt}{0pt}}   % = `bottom' strut

\begin{document}
\title{Measurement of \ce{^{139}La}(p,x) Cross Sections from 35--60 MeV by Stacked-Target Activation}
\author{Jonathan T. Morrell}
\affiliation{Department of Nuclear Engineering, University of California, Berkeley, Berkeley, CA 94720, USA}

\author{Andrew S. Voyles}
\affiliation{Department of Nuclear Engineering, University of California, Berkeley, Berkeley, CA 94720, USA}

\author{M.S. Basunia}
\affiliation{Nuclear Science Division, Lawrence Berkeley National Laboratory, Berkeley, CA 94720, USA}

\author{Jon C. Batchelder}
\affiliation{Department of Nuclear Engineering, University of California, Berkeley, Berkeley, CA 94720, USA}

\author{Eric F. Matthews}
\affiliation{Department of Nuclear Engineering, University of California, Berkeley, Berkeley, CA 94720, USA}

\author{Lee A. Bernstein}
\affiliation{Department of Nuclear Engineering, University of California, Berkeley, Berkeley, CA 94720, USA}
\affiliation{Nuclear Science Division, Lawrence Berkeley National Laboratory, Berkeley, CA 94720, USA}

\date{\today}
\setlength{\parskip}{1em}

\begin{abstract}
A stacked-target of natural lanthanum foils (99.9119\% \ce{^{139}La}) was irradiated using a 60 MeV proton beam at the LBNL 88-Inch Cyclotron.  \ce{^{139}La}(p,x) cross sections are reported between 35--60 MeV for nine product radionuclides.  The primary motivation for this measurement was the need to quantify the production of \ce{^{134}Ce}.  As a positron-emitting analogue of the promising medical radionuclide \ce{^{225}Ac}, \ce{^{134}Ce} is desirable for \textit{in vivo} applications of bio-distribution assays for this emerging radio-pharmaceutical.  The results of this measurement were compared to the nuclear model codes TALYS, EMPIRE and ALICE (using default parameters), which showed significant deviation from the measured values.
\end{abstract}

\maketitle

\section{Introduction}
Proton-induced nuclear reactions in the several tens of MeV incident energy range are commonly used for the production of radionuclides with minimal contaminants, making them a compelling production pathway for diagnostic and therapeutic medical radionuclides \cite{TARKANYI2016262}.  Well-characterized nuclear data for many of these reactions are scarce, yet are critical for researchers wishing to optimize production schemes for these radionuclides \cite{nuclear_data_needs}.

In this work we measured cross sections for the \ce{^{139}La}(p,x) reactions, with a particular interest in the (p,6n) reaction on \ce{^{139}La} (99.9119\% n.a.) for the production of \ce{^{134}Ce}, a medically relevant radionuclide.  In addition, the data from this experiment provide insight into reaction mechanisms and nuclear properties, including pre-eqillibrium particle emission and nuclear level densities, making it useful for benchmarking nuclear reaction modeling codes \cite{OTUKA2014272, PhysRev.87.366, nuclear_data_needs}.

\subsection{Motivation}
Actinium-225 is a promising candidate for new alpha-emitting therapeutic radio-pharmaceuticals \cite{BOLL2005667}.  Actinium-225 has a relatively short half-life of 9.9203 (3) days, and decays to \ce{^{209}Bi} (stable) through the emission of 4 $\alpha$ and 2 $\beta^-$ particles \cite{A225, A221, A217, A213, A209}.  The short range of these $\alpha$ particles is prized for sparing nearby healthy tissue while delivering a lethal dose to the site of disease \cite{Mulford2005}. There are no long-lived products in the decay chain, with the longest activity being the 3.2 h \ce{^{209}Pb} \cite{A209}.
%Also, many potential ligands are available for the chelation of \ce{^{225}Ac}.  
These properties make \ce{^{225}Ac} a very compelling candidate for targeted radionuclide therapy \cite{BOLL2005667}.  In developing biological targeting vectors for the delivery of \ce{^{225}Ac}, their selectivity must be quantified using biodistribution assays \cite{ac225_production}.  The standard for these assays is positron emission tomography (PET), which isn't possible with \ce{^{225}Ac} due to the lack of positron emission in its decay chain.

Instead, \ce{^{134}Ce} has been proposed as a positron emitting analogue of \ce{^{225}Ac}, for potential use in rapid-throughput \textit{in vivo} biodistribution assays.  Cerium-134 decays with a half-life of 3.16 (4) days, which is the closest of the $\beta^+$-emitting Cerium radionuclides to the 10 day half-life of \ce{^{225}Ac} \cite{A134, A225}.  PET imaging of the \ce{^{134}Ce} uptake is performed through the daughter radionuclide \ce{^{134}La}, which $\beta^+$ decays with a short half-life (6.4 minutes) \cite{A134}.  

One proposed mechanism for the production of \ce{^{134}Ce} is through the \ce{^{139}La}(p,6n) reaction, using targets of natural lanthanum, which have a 99.9119\% natural isotopic abundance of \ce{^{139}La}.  This reaction is poorly characterized in the 35--60 MeV energy region, and the predictions of the extensively-used TALYS \cite{TALYS} and EMPIRE \cite{HERMAN20072655} nuclear reaction modeling codes differ by an order of magnitude.  The discrepancies in predictions from these modern codes have profound implications for the design of targets for not only the production of \ce{^{134}Ce}, but also for the production of other radionuclides utilizing energetic proton-induced reactions.  The objective of this paper is to report on a new set of cross section measurements performed using the stacked-target activation technique.  This information will be used to not only quantify the production of \ce{^{134}Ce}, but to gauge the accuracy of several different nuclear reaction models used to predict intermediate light-ion reactions in the energy range below 60 MeV.

\section{Methodology}
In this work we used the stacked-target activation technique, in which one can measure a reaction cross section by quantifying the activity induced within a thin foil of known areal density, using a beam of known intensity \cite{GRAVES201644, VOYLES201853}.  In a single irradiation, many foils can be placed in a ``stack'' (along with monitor foils), yielding cross section measurements at multiple energies by lowering the energy of the primary beam as it traverses the stack.   In this experiment ``degrader'' foils were also included in the stack, to further reduce the beam energy between target foils, such that the measured cross sections fell within the 35--60 MeV range.

%The cross section for a given incident beam energy can be calculated using the activation method following a constant irradiation via the equation
%
%\begin{equation}
%\sigma =  \frac{A_0}{I_p \rho \Delta r (1-e^{-\lambda t_i})}
%\label{eq:xs_calc}
%\end{equation}
%
%\noindent
%where $A_0$ is the end-of-bombardment (EoB) activity of a given reaction product, $I_p$ is the proton beam current, and $\rho \Delta r$ is the areal number density. The factor $(1-e^{-\lambda t_i})$ is a production-decay correction assuming single-step production and decay, where $\lambda$ is the decay constant for a given reaction product and $t_i$ is the total irradiation time.  

%In this experiment the areal number density was determined by measuring the mass $m$, the area $A$ and the thickness $\Delta r$, and was calculated according to the equation
%
%\begin{equation}
%\rho \Delta r=\frac{w\cdot m \cdot N_A}{MA\Delta r}\Delta r = \frac{w\cdot m \cdot N_A}{MA}
%\label{eq:number_density}
%\end{equation}
%
%\noindent
%where $w$ is the isotopic abundance in the sample, $M$ is the molar mass and $N_A$ is Avagadro's number.  

Following irradiation, the end-of-bombardment (EoB) activities of the various proton-induced reaction products were determined by counting the $\gamma$-rays emitted from each foil, using a well-calibrated high-purity germanium (HPGe) detector.  The reaction cross sections (leading to a particular product) were then calculated from the EoB activities.  Many of the produced radionuclides included decay feeding from a parent also produced in the \ce{^{139}La}(p,x) reaction.  Where this contribution was separable, we have reported independent cross sections, however in cases where the parent decay was unable to be measured we have reported cumulative cross sections.

%If a $\gamma$ spectrum is counted for a measurement time $t_m$, beginning some amount of time $t_c$ after the end of irradiation, then the end-of-bombardment activity measured in a photo-peak having $N_c$ counts is given by
%
%\begin{equation}
%A_0 = \frac{\lambda N_c}{(1-e^{-\lambda t_m})e^{-\lambda t_c}I_{\gamma}\epsilon}
%\label{eq:activity}
%\end{equation}
%
%\noindent
%where $I_{\gamma}$ is the $\gamma$ emission fraction per decay and $\epsilon$ is the detector efficiency at the energy of the photo-peak.

The proton beam current incident upon the stack was estimated using a current integrator.  However, the beam current used in the calculation of the cross section was determined more precisely using natural copper and aluminum monitor foils for each lanthanum foil in the stack.  These foils have multiple reaction channels with well-characterized cross sections \cite{IAEACPR}.
%This allows for precise determination of the beam current using the activation technique.

%\begin{equation}
%I_p =  A_0[\sigma \rho \Delta r (1-e^{-\lambda t_i})]^{-1}
%\label{eq:beam_current}
%\end{equation}
%
%Because the proton beam isn't characterized by a single energy, but rather an energy spectrum $\psi(E)$, the cross section used in the beam current calculation should be replaced by an average cross section $\bar{\sigma}$ given by
%
%\begin{equation}
%\bar{\sigma} = \frac{\int_0^{\infty}\sigma(E)\psi(E)dE}{\int_0^{\infty}\psi(E)dE}
%\label{eq:avg_xs}
%\end{equation}

The flux-averaged proton energy associated with each cross section was determined using a Monte Carlo model based on the Anderson \& Ziegler stopping power tables \cite{ZIEGLER20101818}.  This model was optimized to give the best energy assignments using the monitor foil activation measurements, consistent with the technique proposed by Graves et al. in 2016 \cite{GRAVES201644}.  The areal density of each foil in the stack was determined by repeated measurements of the mass and area of each foil.

\subsection{Description of Experiment}

\begin{figure}[htb]
\includegraphics[width=9cm]{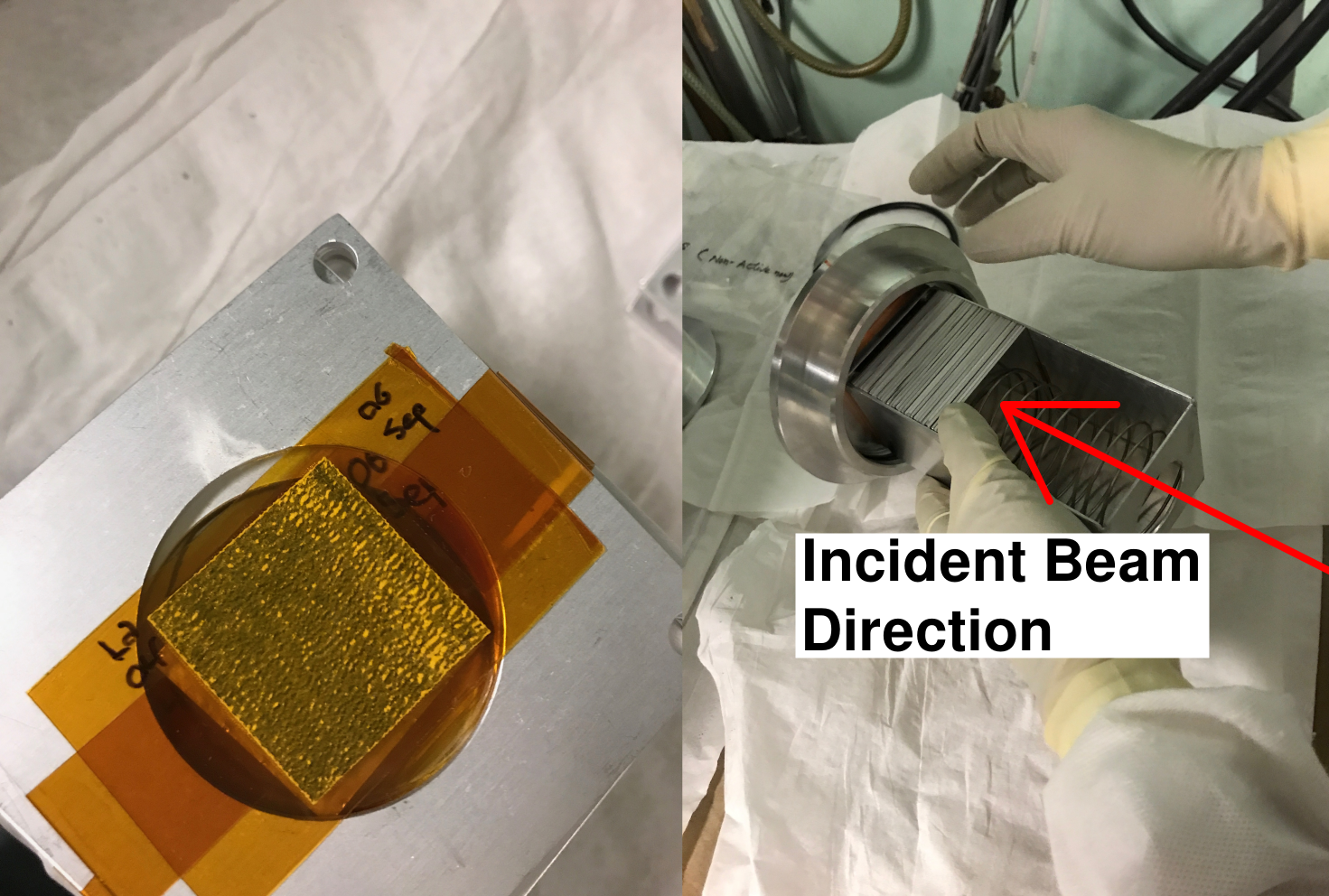}
\caption{Photo of an individual foil packet secured to the aluminum sample holder (left) and the entire foil stack (right).  The front of the stack (facing the beam) is oriented towards the right in this photo.   
}
\label{fig:expt_photos}
\end{figure}

The lanthanum foils used in this experiment were of 99\% purity and were purchased from Goodfellow Corporation (Coraopolis, PA 15108, USA).  The foils were cold rolled to 25 \mmicro m thickness, cut to 1" by 1" squares, and sealed in glass ampules with an inert cover gas (to prevent oxidation).  Just prior to the experiment these ampules were opened, and the foils were removed and cleaned with isopropyl alcohol. The dimensions and masses of these foils were measured, and they were sealed in 3M 5413-Series Kapton polyimide film tape -- each piece of tape consists of 43.2 \mmicro m of a silicone adhesive (nominal 4.79 mg/cm$^2$) on 25.4 \mmicro m of a polyimide backing (nominal 3.61 mg/cm$^2$).  The copper and aluminum monitor foils were cut from 25 \mmicro m-thick sheets into 1" by 1" squares, and were also measured and sealed in Kapton tape. These foil packets were then secured over the hollow aperture of 2.25" by 2.25" aluminum sample holders (see Fig. \ref{fig:expt_photos}), which protected the foils during handling and centered them in the beam pipe.  Ten sets, each consisting of a single aluminum, copper and lanthanum foil packet, were prepared in this manner for cross section measurements at ten different energies.

Multiple plates of 6061-aluminum, also 2.25" by 2.25", were placed in between each set of foil packets to degrade the beam energy by a few MeV between each foil packet, allowing cross section measurements over a range of proton energies from 35--60 MeV.
%The appropriate thickness for each aluminum degrader was determined using an Anderson \& Ziegler calculation, however the actual proton energy distributions seen by each foil were more carefully calculated using the monitor foil reactions.  This is described in more detail in section \ref{monitors}.
Additionally, stainless steel plates (approximately 100 mg/cm$^2$) were placed at the front and back of the stack.  Post-irradiation dose mapping of the activated stainless plates using radiochromic film (Gafchromic EBT3) confirmed that the proton beam was centered on the samples, and that the entirety of the $\approx$1 cm-diameter primary proton beam was contained well within the 1" by 1" borders of the foil packets.  This method is consistent with previously established techniques \cite{GRAVES201644, VOYLES201853}.

A single ORTEC GMX Series (model GMX-50220-S) High-Purity Germanium (HPGe) detector was used in this experiment. The detector is a nitrogen-cooled coaxial n-type HPGe with a 0.5 mm beryllium window, and a 64.9 mm diameter, 57.8 mm long crystal.  The energy and photopeak efficiency of the HPGe detector used in this measurement were calibrated using four standard calibration sources of known activity (rel. error $<$1\%): \ce{^{137}Cs}, \ce{^{152}Eu}, \ce{^{54}Mn} and \ce{^{133}Ba}.  The photopeak efficiencies were also corrected for detector dead-time, as well as self-attenuation within the foils --- using photon attenuation cross sections retrieved from the XCOM database \cite{berger2011xcom}.

\subsection{Facility Overview}
This experiment took place using a proton beam at the 88-Inch Cyclotron located at Lawrence Berkeley National Laboratory (LBNL) in Berkeley, California \cite{One_Stop_Shop}.  The 88-Inch Cyclotron is a variable-beam, variable-energy K=140 isochronous cyclotron, with a maximum recorded proton energy of 60 MeV and a maximum proton beam current of approximately 20 \mmicro A.

\begin{figure}[htb]
\includegraphics[width=9cm]{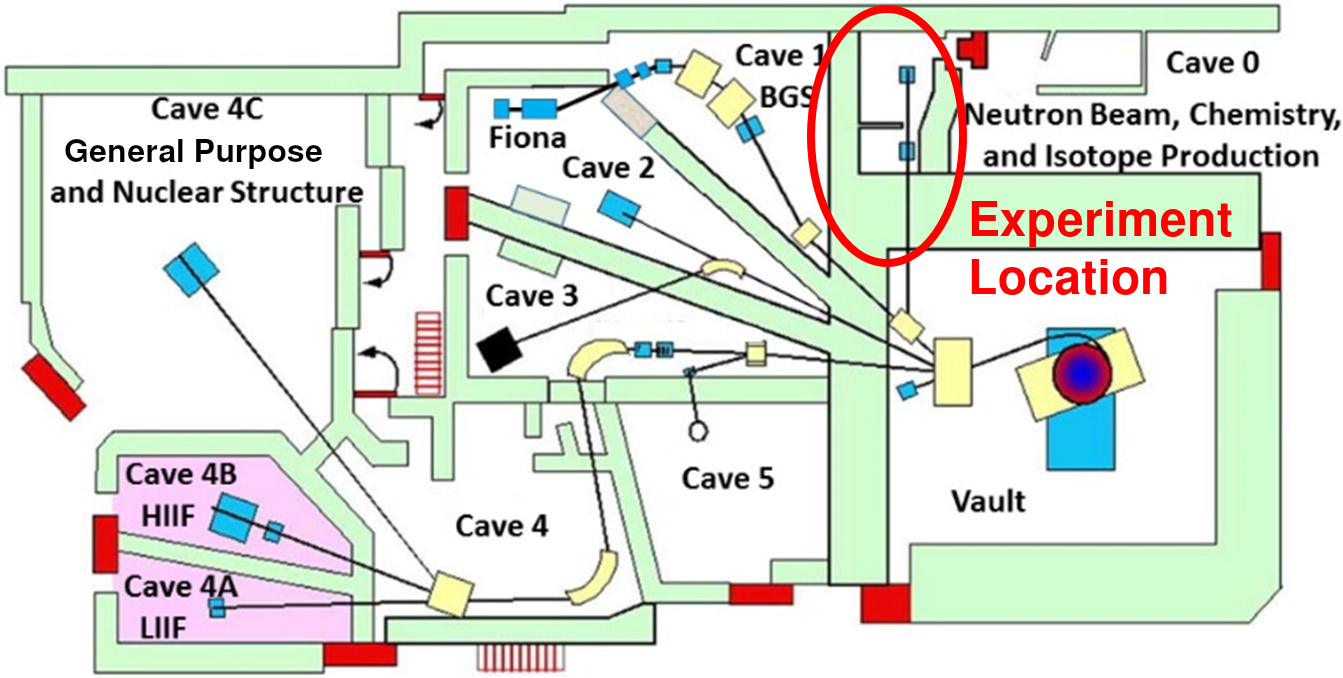}
\caption{Graphical representation of the 88-Inch Cyclotron floor plan.  The irradiation described here took place in Cave 0, which is circled above in red.
}
\label{fig:cyclotron}
\end{figure}

The 88-Inch Cyclotron facility has several isolated beamlines for a multitude of applications (see Fig. \ref{fig:cyclotron}).  This experiment took place in Cave 0, which has a $\sim$3 m beamline that is shielded from any neutron radiation produced in the cyclotron vault.  The target holder for the foil stack was mounted at the end of this beamline, downstream from two bending magnets and several focusing quadrupoles in the main cyclotron vault.

This experiment marked the first time in recent history that the 88-Inch Cyclotron had attempted extracting a 60 MeV proton beam.  Due to RF-power and beam optics limitations, the result was extremely low transmission with only approximately 8 nA of beam current  ($\leq$0.1\% transmission efficiency).  A subsequent tune to Cave 0 exhibited significantly higher transmission, suggesting that better performance can be expected in future energetic proton runs.  It was later determined by the monitor foil activation that the mean beam energy was approximately 57 MeV, and not 60 MeV as originally desired, further indicating that the initial tune was far from optimized for 60 MeV protons.

\subsection{Irradiation and Counting}
The foils were irradiated with 8 nA of proton beam current for 1 hour, 37 minutes and 24 seconds.  The total collected charge of the beam was measured using a current integrator connected to the electrically-isolated target holder, which was used to determine that the beam current was stable over the duration of the experiment.  This measurement of the beam current incident upon the target holder was also used to validate the beam current values determined using the monitor foil activation.

After irradiation, the foils were removed from the beamline and transferred to the HPGe counting lab approximately 15 minutes after EoB.  Upon removal it was discovered that the third lanthanum foil showed excessive oxidation and had ruptured its Kapton encapsulation (prior to counting), indicating potential material loss. Therefore, in the interest of surety, no cross sections will be reported for this foil.

The foils were counted for four weeks following EoB.  Each foil was counted multiple times, in order to reduce uncertainty and aid in isotope identification.  This lengthy counting duration was necessary because the 604.6 and 606.8 keV $\gamma$-rays from the \ce{^{135}Ce} isotope significantly contaminated the 604.7 keV line emanating from the \ce{^{134}Ce} daughter isotope \ce{^{134}La} \cite{A135, A134}. Because the \ce{^{134}Ce} isotope has a longer half-life (3.16 days vs 17.7 hours), the 604.7 keV $\gamma$-ray was able to be resolved after the \ce{^{135}Ce} isotope had decayed to negligible levels \cite{A135, A134}.

\section{Data Analysis}
The general procedure for calculating cross sections proceeded as follows.  First, every $\gamma$ line emitted from each isotope of interest was fit in each spectrum collected.  The number of counts in each peak was used to determine the activity of the isotope at the time the spectrum was taken.  These activities as a function of ``cooling'' time (time since EoB) were used to calculate the EoB activity, $A_0$, for that isotope.  Each $A_0$ was then used to determine a cross section (for the lanthanum foil data) or a beam current (for the copper/aluminum foil data).  The energy assignments for each foil were determined using the variance minimization approach proposed by Graves \cite{GRAVES201644}, as discussed in section \ref{monitors}.  The NPAT code, developed at UC Berkeley \cite{npat}, was used for spectrum analysis, fitting decay curves, and calculating the proton energy spectrum in each foil.

The uncertainties in the reported cross sections had five main contributions: uncertainties in evaluated half-lives and gamma intensities ($\approx$1\%), EoB activity determination ($\approx$1\%), detector efficiency calibration ($\approx$3\%), foil areal density ($\approx$1\%) and proton current determination ($\approx$5\%).  Each contribution to the uncertainty was assumed to be independent and was added in quadrature.

\subsection{Peak Fitting}

\begin{figure}[htb]
\includegraphics[width=9cm]{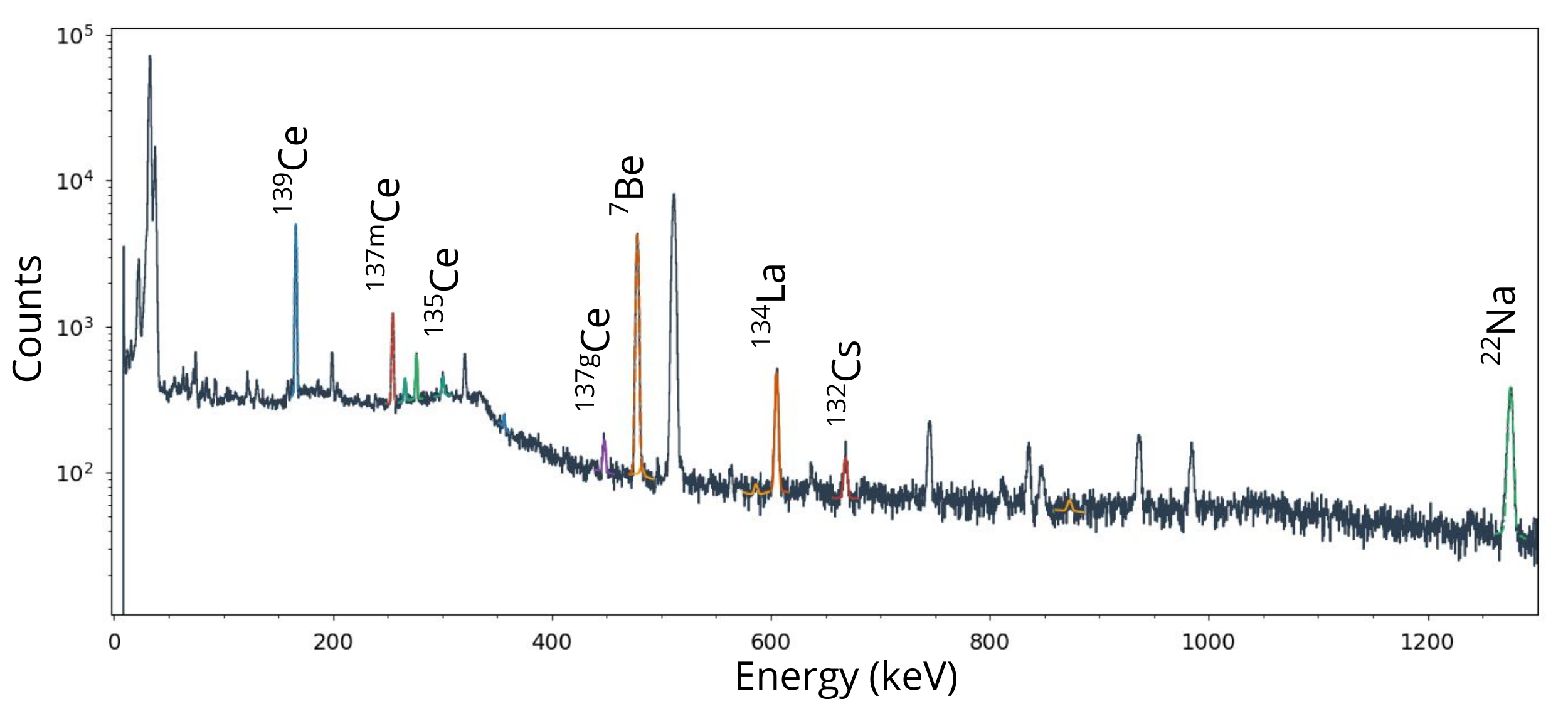}
\caption{A $\gamma$-ray spectrum collected from the lanthanum foil activated with approximately 56 MeV protons.
}
\label{fig:peak_fit}
\end{figure}

The detector energy and efficiency calibration, as well as the induced activity in each sample, were determined by peak fitting to the individual spectra.  Energy centroids and relative intensities were constrained with some uncertainty by the decay data from ENSDF \cite{ensdf}, also listed in Appendix \ref{nudat_appendix}.  Each peak was fit with a skewed Gaussian function on top of a linear background \cite{Knoll}. As implemented in the NPAT code \cite{npat}, the complete functional form of the peak fit, $F(i)$, as a function of channel number $i$ is as follows.

\begin{multline}
F_{peak}(i) = m\cdot i + b + A\cdot [\exp (-\frac{(i-\mu)^2}{2\sigma^2}) \\
 + R\cdot \exp (\frac{i-\mu}{\alpha \sigma}) \erfc (\frac{i-\mu}{\sqrt{2}\sigma}+\frac{1}{\sqrt{2}\alpha})]
\label{eq:peak}
\end{multline}

\noindent
where $m\cdot i+b$ is the background component, $A\cdot \exp (-\frac{(i-\mu)^2}{2\sigma^2})$ is the (dominant) Gaussian component, and $A\cdot R\cdot \exp (\frac{i-\mu}{\alpha \sigma}) \erfc (\frac{i-\mu}{\sqrt{2}\sigma}+\frac{1}{\sqrt{2}\alpha})$ is the ``tailing'' component.  Typical values of $R$ and $\alpha$ are $\approx$0.2 and $\approx$0.9, respectively.

The number of counts in a photopeak fit using this functional form is given by

\begin{equation}
N_c = A \big(\sqrt{2\pi} \sigma + 2R \alpha \sigma \exp (-\frac{1}{2\alpha^2})\big)
\label{eq:counts}
\end{equation}

An example of a measured $\gamma$-ray spectrum is shown in Fig. \ref{fig:peak_fit}, with photopeak fits superimposed on the spectrum.

\subsection{Determining Foil Activities}
\label{eob_activities}

To obtain the EoB activities for each proton-induced reaction product, in each foil, we determine the apparent activity from each photopeak that was observed and perform a fit to the decay curve generated from the appropriate Bateman equations \cite{Bateman}.

For a single photopeak having $N_c$ counts observed with efficiency $\epsilon$ from a radionuclide with decay constant $\lambda$ and intensity $I_{\gamma}$, the apparent activity in a photopeak at some cooling time $t_c$ after the end-of-bombardment is given by by

\begin{equation}
A(t_c) = \frac{\lambda N_c}{(1-e^{-\lambda t_m})I_{\gamma}\epsilon}
\end{equation}

\noindent
where $t_m$ is the measurement time.  If the population of this nucleus has no contribution from the decay of a parent, the activity $A_0$ can be determined using a fit to the equation

\begin{equation}
A(t_c) = A_0e^{-\lambda t_c}
\label{eq:decay}
\end{equation}

\begin{figure}[htb]
\includegraphics[width=9cm]{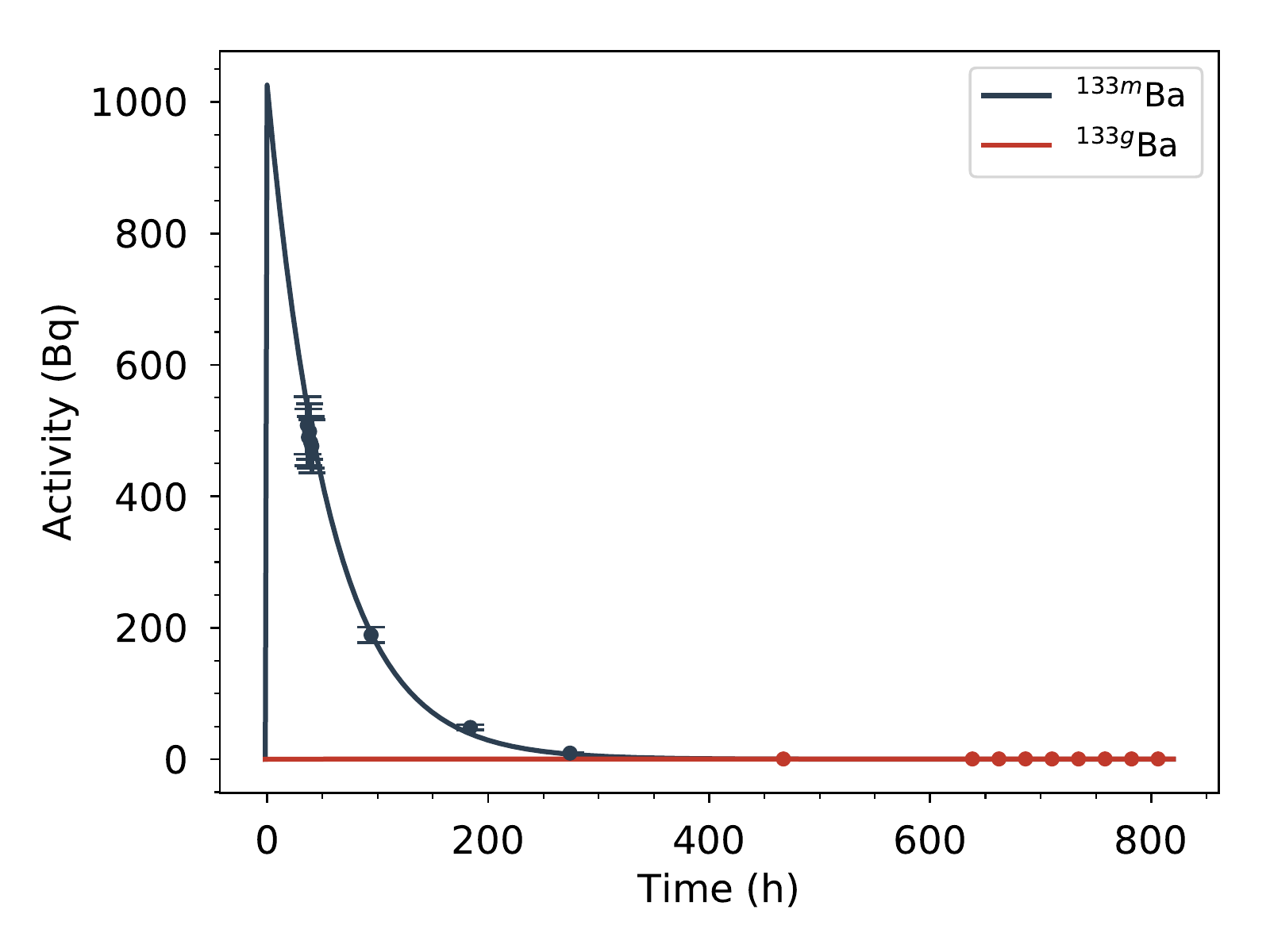}
\caption{Example of a decay curve and associated exponential fit used to calculate the initial activity of the \ce{^{133m}Ba} and \ce{^{133g}Ba} isotopes in the 1$^{st}$ lanthanum foil.  The uncertainty in the activity was dominated by counting statistics and the evaluated half-life for most of the observed reaction products.
}
\label{fig:decay_curves}
\end{figure}

Eq. \ref{eq:decay} is only valid for single-step decay pathways, and several activation products in this experiment exhibit multi-step decays.  For example, \ce{^{134}Ce} decays to \ce{^{134}La} which then decays to the stable \ce{^{134}Ba} \cite{A134}.  For these two-step decay chains, the decay curve will take the form

\begin{equation}
A_D(t_c) = A_{P_0}R_b\frac{\lambda_D}{\lambda_D - \lambda_P}(e^{-\lambda_P t_c}-e^{-\lambda_D t_c})+A_{D_0}e^{-\lambda_D t_c}
\end{equation}

\noindent
where $R_b$ is the branching-ratio, and the subscripts $P$ and $D$ indicate the parent and daughter isotopes respectively. An example of a fit to this exponential decay curve is shown in Fig. \ref{fig:decay_curves}.  This calculation of the EoB activities required a measurement of the initial parent activity $A_{P_0}$, again using a fit to Eq. \ref{eq:decay}, which somewhat increased the uncertainty in the calculation of the initial daughter activities.  However, most EoB activities were still quantified to approximately 1--3\% relative uncertainty.

\subsection{Current Monitors and Energy Assignments}
\label{monitors}

%The proton beam current incident upon each foil can be calculated for a given monitor reaction channel according to Eq. \ref{eq:beam_current}, where $A_0$ is determined by activation measurements as previously described. 

Using the end-of-bombardment activities $A_0$ determined by the activation spectra and the measured areal densities $\rho r$ of each foil, we can calculate the proton beam current (in units of protons per second) $I_p$ incident upon each monitor foil according to

\begin{equation}
I_p = \frac{A_0}{\bar{\sigma}(\rho r)(1-e^{-\lambda t_i})}
\label{eq:beam_current}
\end{equation}

\noindent
where the factor $(1-e^{-\lambda t_i})$ accounts for decay during a constant production interval $t_i$, and the flux-weighted cross section $\bar{\sigma}$ is given by

\begin{equation}
\bar{\sigma} = \frac{\int_0^{\infty}\sigma(E)\psi(E)dE}{\int_0^{\infty}\psi(E)dE}
\label{eq:avg_xs}
\end{equation}

\noindent
where $\sigma(E)$ comes from the IAEA-recommended cross sections \cite{Hermanne2018} and $\psi(E)$ is the energy spectrum of the proton flux.

\begin{figure}[htb]
\includegraphics[width=9cm]{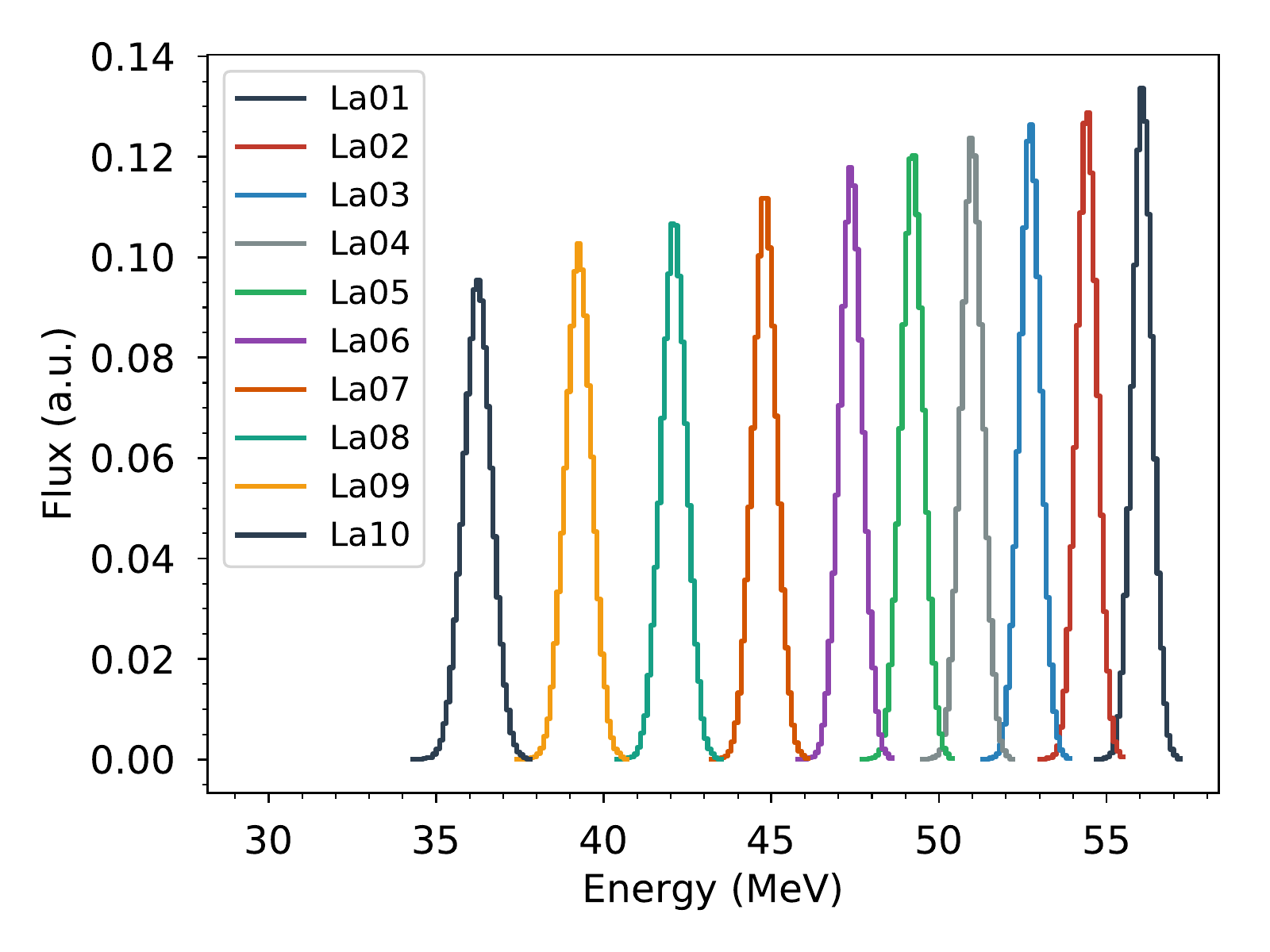}
\caption{Plot of the calculated proton energy spectra for each lanthanum foil in the target stack.
}
\label{fig:az_spectrum}
\end{figure}

This treatment accounts for the fact that the beam has a finite energy width that increases toward the back of the stack due to energy straggling of the beam.  The proton flux spectrum $\psi(E)$ was determined using an Anderson \& Ziegler-based Monte Carlo code, as implemented in NPAT \cite{npat, ZIEGLER20101818}.  A plot showing the flux spectra for each foil in the stack, as predicted by the Anderson \& Ziegler-based model, is shown in Fig. \ref{fig:az_spectrum}.

\begin{figure}[htb]
\includegraphics[width=9cm]{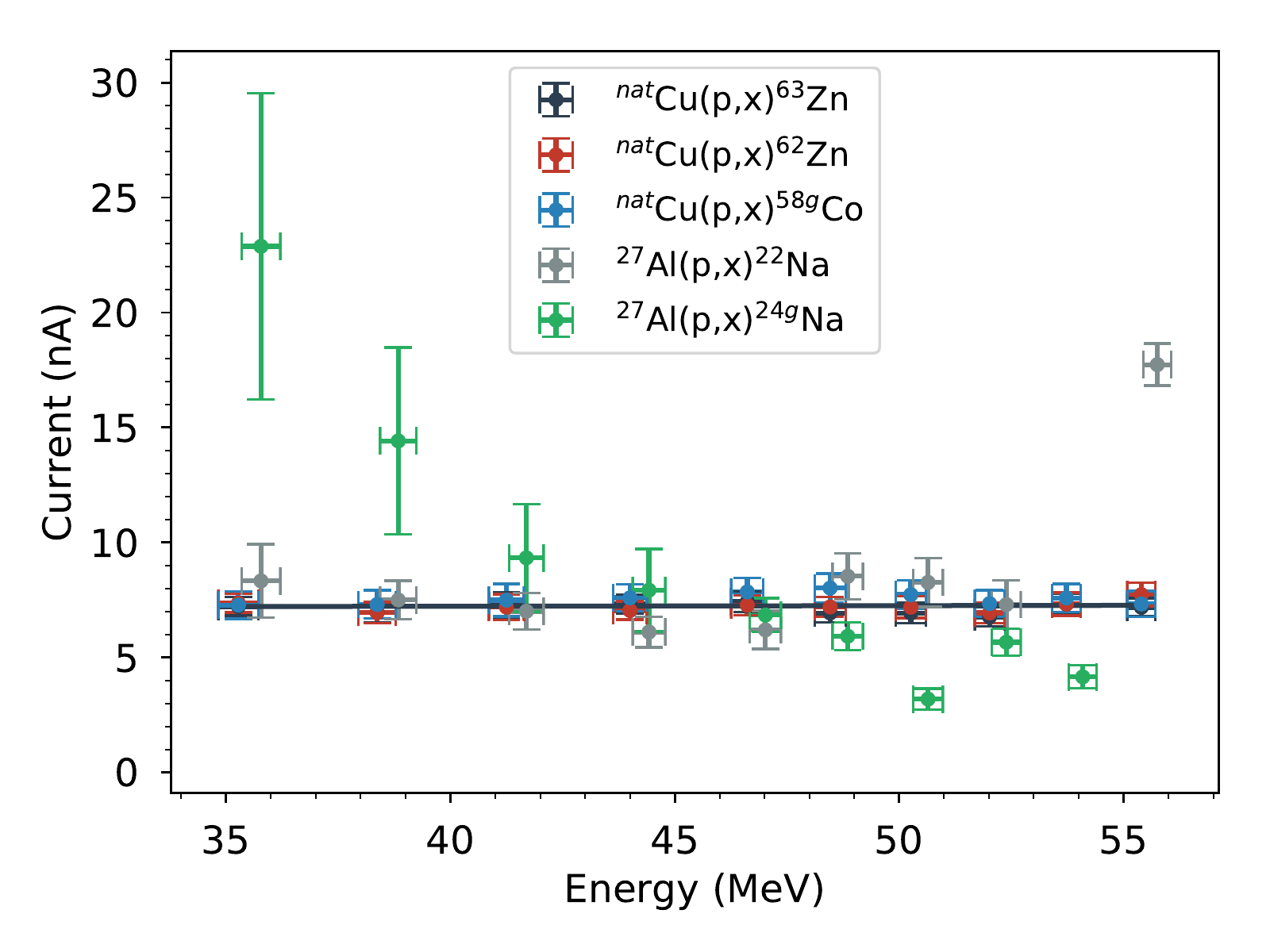}
\caption{Plot of the proton beam current measured by each of the monitor foil reaction channels, along with a linear fit that was used to calculate the current for the lanthanum foils.  The aluminum monitor channels are plotted only to illustrate the magnitude of the uncertainty due to contaminating reactions in the Kapton, and were not included in the final analysis.
}
\label{fig:beam_current}
\end{figure}

Measured currents for the three copper monitor reactions and the two aluminum reactions are plotted in Fig. \ref{fig:beam_current}, as well as a linear fit to these values, which was used to interpolate the beam current witnessed by the lanthanum foils.  One would expect the proton beam current to decrease as it traverses the stack as the beam reacts, scatters and diffuses out of the path of downstream foils.  However, only a very small decrease in beam current was observed; therefore, a linear fit proved sufficient for the interpolation.

As can be seen in Fig. \ref{fig:beam_current}, the uncertainty in both aluminum monitor reactions was exceptionally large, particularly in the \ce{^{27}Al}(p,x)\ce{^{22}Na} reaction.  This was due to corrections for contaminating reactions in the silicone-based adhesive of the Kapton tape that sealed the foils (\ce{^{28}Si}(p,x)\ce{^{22}Na} and \ce{^{28}Si}(p,x)\ce{^{24}Na}), as well as reactions on the aluminum frames.  These corrections were performed by measuring the activities of the \ce{^{22}Na} and \ce{^{24}Na} isotopes in the Kapton sealing the neighboring lanthanum and copper foils, and subtracting this contribution from the aluminum foil data.

Secondary neutron production could have also contributed to \ce{^{22,24}Na} activation.  However, the secondary neutron flux predicted by an MCNP model of the experiment was 3--4 orders of magnitude lower than the proton flux, suggesting that it has a minor effect on the measured monitor reaction activities \cite{MCNP}.

To prevent systematic errors, it was decided to remove the aluminum monitor channels from the analysis, however the (post-correction) results from these channels are still plotted in Fig. \ref{fig:beam_current}, to illustrate the magnitude of the error induced by contaminating reactions in the aluminum frames and silicone adhesive.

\subsection{Optimization of Energy Assignments}

The flux-averaged proton energies and 1$\sigma$-widths of the proton energy distributions for each foil were first estimated using the Anderson \& Ziegler formalism for proton transport (stack design listed in Appendix B).  The apparent proton current in each monitor foil was measured using the IAEA-recommended cross sections for the \ce{^{nat}Al}(p,x)\ce{^{22}Na}, \ce{^{nat}Al}(p,x)\ce{^{24}Na}, \ce{^{nat}Cu}(p,x)\ce{^{62}Zn}, \ce{^{nat}Cu}(p,x)\ce{^{63}Zn}, and \ce{^{nat}Cu}(p,x)\ce{^{58}Co} monitor reactions \cite{Hermanne2018}. Disagreement in the apparent beam current between monitor channels was observed, particularly for the foils on the low-energy side of the stack, which was where the energy dependence of the monitor cross sections varied most strongly.  This was indicative of incorrect characterization of the proton energy spectra incident on each monitor foil.

To correct this discrepancy, the effective density of the 6061-aluminum degrader foils was treated as a free parameter in the energy loss calculation, and was optimized with respect to the reduced $\chi^2$ of the (linear) fit to the monitor reaction data.  The minimum value of $\chi^2_{\nu}$ yields the energy assignments which give the best agreement between the proton current derived from the various monitor reaction channels.  This ``variance minimization'' technique was performed in a manner consistent with Graves (2016) and Voyles (2018) \cite{GRAVES201644, VOYLES201853}.  Afterwards, because the discrepancy was still quite large for foils in the high-energy portion of the stack ($\approx$30\%), the incident beam energy was also treated as a free parameter and this same minimization was performed on density and incident energy simultaneously. This resulted in a more reasonable density change of -2\%, for an average incident beam energy of 57 MeV.

It should be noted that this variance minimization approach does not necessarily imply that the degrader density was physically less or greater than was measured, but instead serves as a correction for stopping power characterization and enhanced energy loss due to unaccounted systematics in the original stack design.  In essence, this method assigns energy spectra to each foil which best match the shape of the apparent monitor reaction cross sections to the shape of the IAEA-recommended charged-particle reference standards.

\begin{figure}[htb]
\includegraphics[width=9cm]{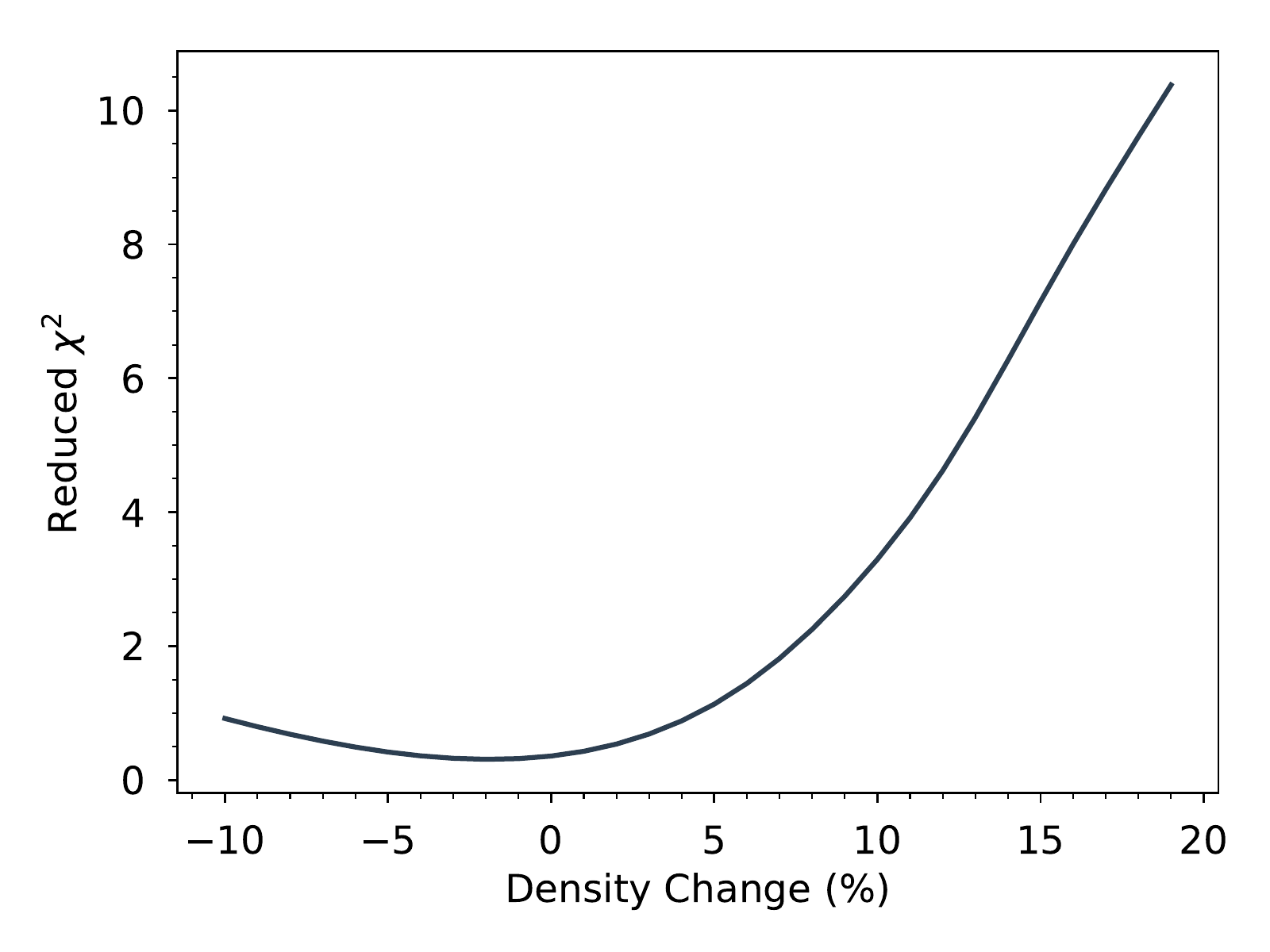}
\caption{Plot of the reduced $\chi^2$ figure-of-merit for the current monitor data, as the effective density of the degraders was varied.
}
\label{fig:minimization}
\end{figure}

Fig. \ref{fig:minimization} shows the results of this minimization for an incident proton energy of 57 MeV.  The reduced $\chi^2$ of the linear fit to the monitor foil currents (shown in Fig. \ref{fig:beam_current}) was used to determine the optimum energy assignments, based on the Anderson \& Ziegler proton transport model, by varying the effective areal density of the 6061-aluminum degraders.

This minimization shows that the optimum energy assignments result from a -2\% change in the effective areal density of the stack. Additionally, it implies that the average incident proton beam energy was 57 MeV, rather than the expected 60 MeV, an unexpected deviation from the initial estimates.  This discrepancy could be attributable to a number of experimental factors.  This was the highest-energy proton beam in recent history at the LBNL 88-Inch Cyclotron, and the cyclotron tuning solutions were observed to be significantly different than for beams run more frequently at the facility.  This is supported by the fact the transmission out of the cyclotron was very low, about 0.1\%, and that a subsequent retune of the machine (after the irradiation) yielded a much better transmission. 

%As for the 15\% deviation from the expected areal density, this could be attributed to a measurement error (of the foil densities), or a modeling error.  Interestingly, an Anderson and Ziegler (SRIM/TRIM) calculation gave optimum energy assignments for 0\% deviation, which means that something may have been wrong with either the MCNP input, or the data it uses.

This monitor foil variance minimization technique was also performed with the Monte Carlo code MCNP \cite{MCNP}.  This corresponded to a 15\% enhancement in the degrader areal density, which is significantly higher than comparable values in the literature \cite{GRAVES201644, VOYLES201853}. These results suggest a systematic issue in the low-energy charged-particle stopping power tables used by MCNP, and that a detailed comparison between the Anderson \& Ziegler and MCNP stopping powers should be further explored in this intermediate energy range.

\section{Results and Discussion}

Using the end-of-bombardment activities, beam currents, and energy assignments determined in the previous section, the flux-averaged cross sections were calculated with the following equation

\begin{equation}
\sigma = \frac{A_0}{I_p(\rho r)(1-e^{-\lambda t_i})}
\end{equation}

The results of these cross section measurements are summarized in Table \ref{tab:cross_sections}, and are described in detail below.

These results were compared to the TENDL-2017 evaluation and predictions from the TALYS-1.9, EMPIRE-3.2 and ALICE-20 nuclear reaction modeling codes, all  using default parameters \cite{HERMAN20072655, TALYS, ALICE}.  The pre-equilibrium model used in the EMPIRE and ALICE calculations was the Hybrid Monte-Carlo Simulation module (HMS), while the TALYS code uses an exciton pre-equilibrium model \cite{HERMAN20072655, TALYS, ALICE}.  Many of the excitation functions had a characteristic ``compound peak'' corresponding to energies between the threshold and the opening of the next significantly populated exit channel.

These measurements were also compared to the work of T{\'a}rk{\'a}nyi et al., who performed a similar stacked target measurement on \ce{^{nat}La} in 2017 \cite{Tarkanyi2017}.  There were some discrepancies between the results of this work and the T{\'a}rk{\'a}nyi measurements, however there wasn't a clear systematic bias between one set of measurements and the other.  Because the largest discrepancies were observed in channels featuring multi-step decay, it is most likely that differences in counting schedules and multi-step decay fits were the major sources of systematic discrepancies between the two experiments.

\begin{table*}
\begin{ruledtabular}
\begin{tabular}{cccccccccc}
 & \multicolumn{9}{c}{\ce{^{139}La}(p,x) Production cross section (mb)} \Bstrut \\
\hline
\Tstrut\Bstrut E$_p$ (MeV) \\
 & 56.07(31) & 54.42(32) & 50.99(33) & 49.21(34) & 47.38(35) & 44.80(37) & 42.09(39) & 39.25(41) & 36.23(44) \Bstrut \\
\hline
\Tstrut \ce{^{134}Ce}$_{i}$ \\
 & 50.7(88) & 20.3(97) & 3.9(25) & 2.55(38) & - & - & - & - & - \\
\ce{^{135}Ce}$_{c}$ \\
 & 458(23) & 474(23) & 377(17) & 332(15) & 254(14) & 123.4(30) & 29.00(82) & 2.90(16) & 0.493(37) \\
\ce{^{137m}Ce}$_{i}$ \\
 & 79.9(47) & 89.2(51) & 95.9(52) & 112.4(59) & 124.5(75) & 148.3(46) & 179.9(52) & 306(14) & 426(14) \\
\ce{^{137g}Ce}$_{i}$ \\
 & 23.5(51) & 26.2(57) & 27.4(68) & 32.3(80) & 37.4(74) & 40.8(78) & 48.7(96) & 77(18) & 114(30) \\
\ce{^{139}Ce}$_{c}$ \\
 & 18.9(12) & 19.1(19) & 20.8(17) & 17.6(16) & 19.8(17) & 23.9(11) & 23.29(78) & 31.7(16) & 29.9(11) \\
\ce{^{132}Cs}$_{i}$ \\
 & 0.170(14) & 0.1114(54) & 0.0405(57) & - & - & - & - & - & - \\
\ce{^{133m}Ba}$_{i}$ \\
 & 12.54(72) & 13.89(79) & 13.74(74) & 13.76(69) & 12.57(77) & 9.21(30) & 5.17(15) & 2.67(24) & 0.682(81) \\
\ce{^{133g}Ba}$_{i}$ \\
 & 2.82(67) & 3.1(16) & 2.6(13) & 5.0(11) & 3.4(12) & 3.87(93) & 1.77(58) & - & - \\
\ce{^{135}La}$_{i}$ \\
 & 191(61) & 176(77) & 77(38) & 77(39) & 80(57) & - & - & - & - \Bstrut \\
\\
\hline
\hline
 & \multicolumn{8}{c}{\ce{^{nat}Cu}(p,x) Production cross section (mb)} \Tstrut\Bstrut \\
\hline
\Tstrut E$_p$ (MeV) \\
 & 55.40(32) & 53.73(33) & 52.02(34) & 50.26(34) & 48.46(35) & 46.62(36) & 44.00(38) & 41.25(40) & 38.37(42) \\
 & 35.28(45) \Bstrut \\
\hline
\Tstrut \ce{^{61}Cu}$_{c}$ \\
 & 83.2(27) & 89.7(33) & 93.0(18) & 101.1(31) & 110.3(37) & 120.3(35) & 142.4(20) & 164.7(93) & 183.3(72) \\
 & 182.7(44) \\
\end{tabular}
\label{table:wide_XS}
\end{ruledtabular}
\caption{\label{tab:cross_sections} Summary of cross sections measured in this work.  Subsripts \emph{c} and \emph{i} indicate cumulative and independant cross sections, respectively.}
\end{table*}
\ \

\subsection{\ce{^{139}La}(p,6n)\ce{^{134}Ce} Cross Section}

\ce{^{134}Ce} undergoes electron capture decay to the \ce{^{134}La} ground state with a 98.9\% branching ratio, with a 0.209\%, 130.4 keV $\gamma$-ray being the strongest line \cite{A134}.  The low-intensity and energy of this transition led us to choose the 604.721 keV ($I_{\gamma}$=5.04\%) line in the decay of the daughter isotope \ce{^{134}La} to measure the \ce{^{139}La}(p,6n)\ce{^{134}Ce} cross section.  Because \ce{^{134}La} has a 6.45 minute half-life, all 604.7 keV $\gamma$'s measured after several hours of decay time were attributable only to the decay of the initial \ce{^{134}Ce} population, regardless of the initial population of \ce{^{134}La}.

An additional complication was that \ce{^{135}Ce} produces multiple decay $\gamma$'s very close in energy to the 604.7 keV line. Because of the 17.7 (3) hour half-life of \ce{^{135}Ce}, roughly two weeks of post-irradiation decay time were required to measure the \ce{^{134}Ce} activity without contaminating $\gamma$ lines from \ce{^{135}Ce}.

\begin{figure}[htb]
\includegraphics[width=9cm]{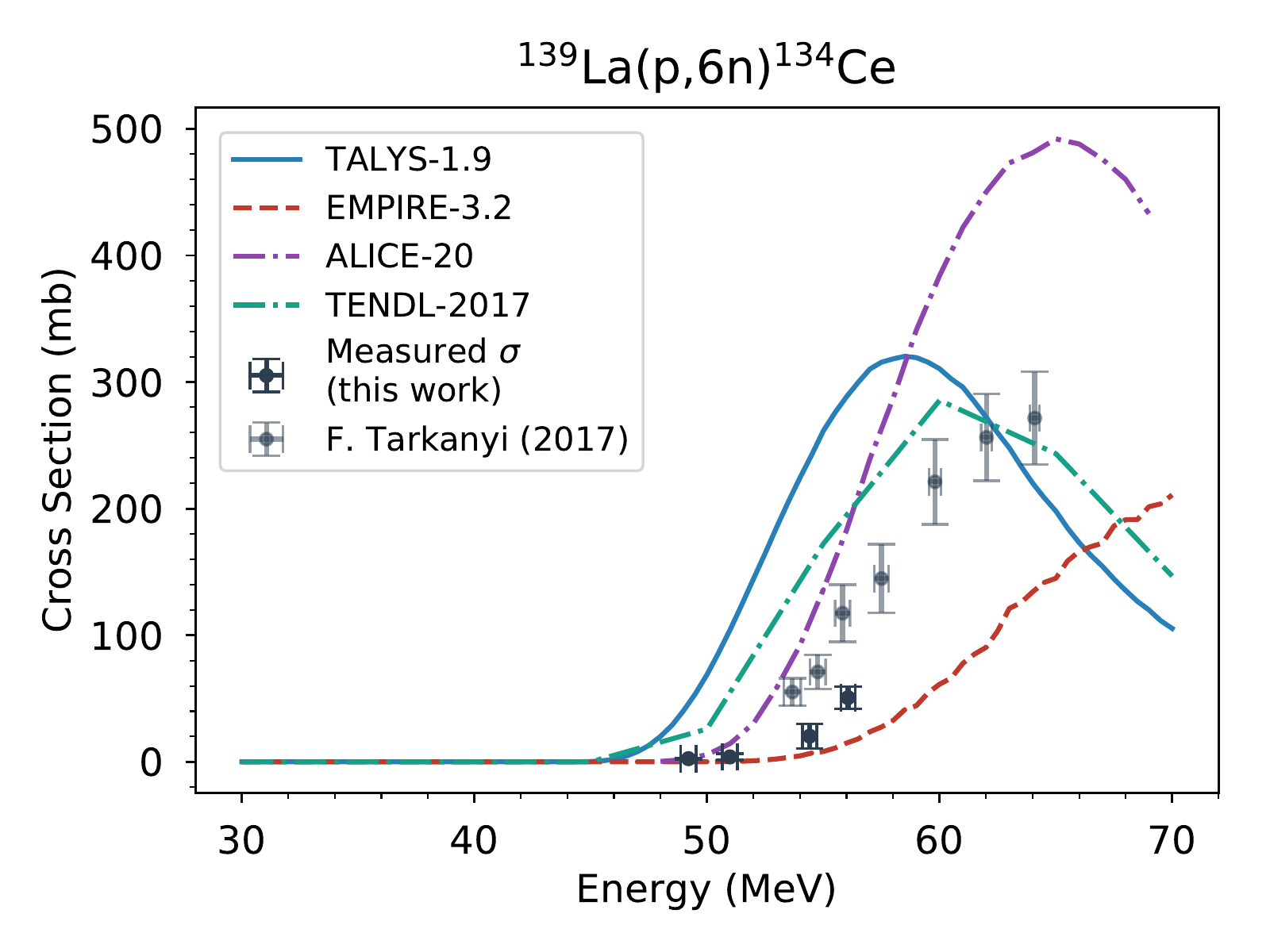}
\caption{Measured cross sections for the \ce{^{139}La}(p,6n)\ce{^{134}Ce} reaction.
}
\label{fig:134CE}
\end{figure}

The measured cross sections for the production of \ce{^{134}Ce} are plotted in Fig. \ref{fig:134CE}.  The hybrid Monte Carlo simulation (HMS) pre-equilibrium model used by EMPIRE seems to slightly over-estimate the centroid energy of the ``compound peak'' seen in the measured data, whereas the exciton model used by TALYS underestimates the peak in the (p,6n) channel by 5--10 MeV.  ALICE also uses a HMS model, and while it accurately estimates the centroid of the cross section it significantly overestimates the magnitude.  This was not necessarily the case for all the measured reaction channels, but a similar trend could be seen for the (p,5n) channel as well (Fig. \ref{fig:135CE}).

This was likely attributable to differences in the pre-equilibrium models between the codes.  Because particles emitted in pre-equilibrium carry a significant amount of energy out of the nucleus before compound nucleus formation, small differences in these models can greatly affect which compound nucleus is formed at a given incident proton energy, shifting the centroid energy of the ``compound peak''.  And while the TENDL-2017 evaluation (based on TALYS-1.9 \cite{TALYS}) better matches data than the modeling codes, it still underestimates the energy of the peak in the cross section by about 5 MeV.

%was likely attributable to the fact that charged particle data in the pre-equilibrium energy range is scarce, so the default parameters used in the model wouldn't be expected to give reliable results without some adjustment.

\subsection{\ce{^{139}La}(p,5n)\ce{^{135}Ce} Cross Section}

The \ce{^{139}La}(p,5n)\ce{^{135}Ce} reaction was perhaps the most accurately quantified, due to a high number of intense $\gamma$ emissions (e.g. 41.8\% for the 265.56 keV line) and a 17.7 hour half-life.  Because the \ce{^{135m}Ce} isomer (t$_{1/2}$=20 s) had completely decayed by the time the foils had been transferred to the counting lab, the reported cross sections for this reaction are cumulative.  %The uncertainties were on the order of 6\%, of which roughly 3\% came from the $I_{\gamma}$'s given by Nuclear Data Sheets \cite{A135} and 1\% from the uncertainty in half-life.  

\begin{figure}[htb]
\includegraphics[width=9cm]{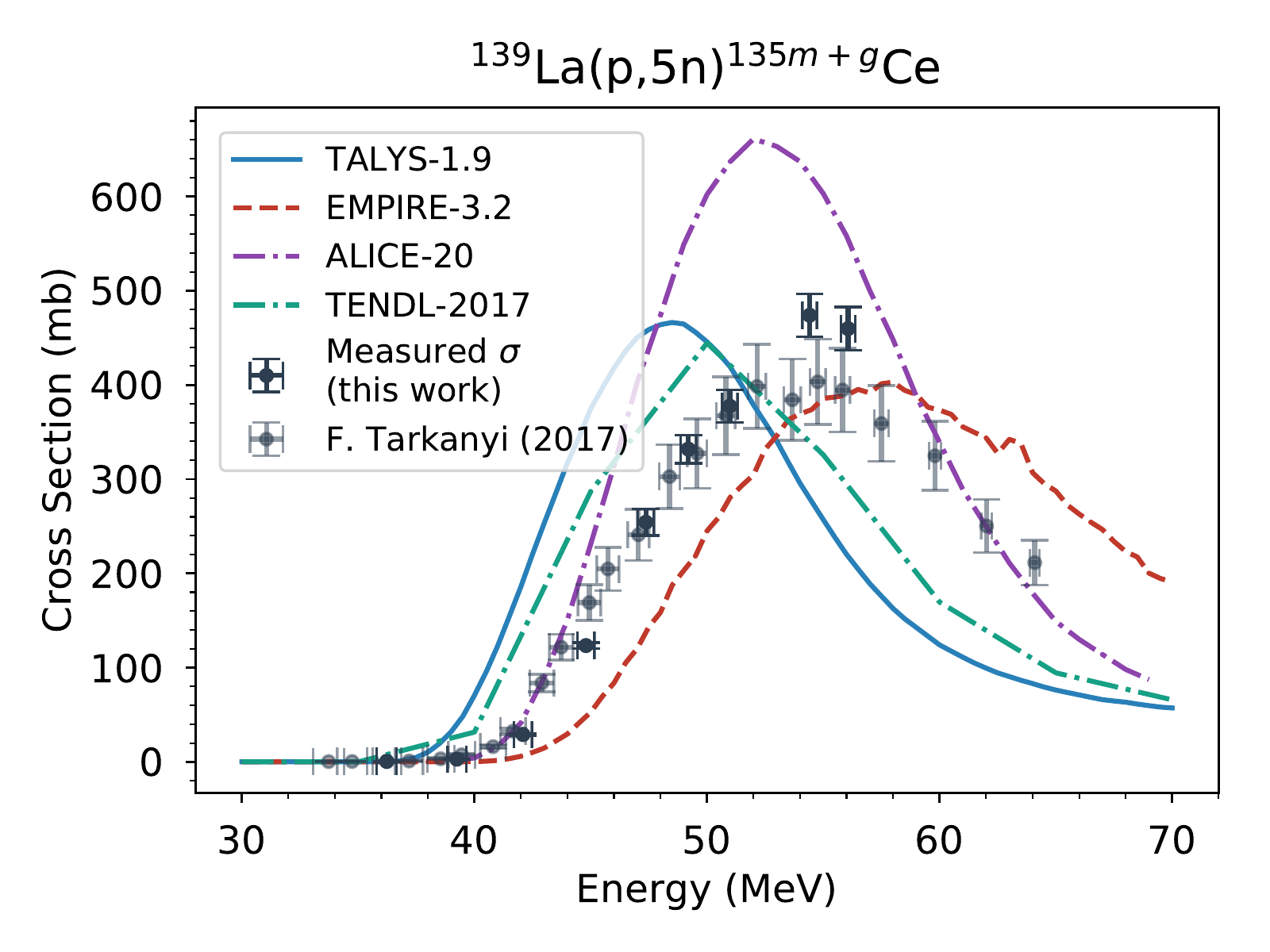}
\caption{Measured cross sections for the \ce{^{139}La}(p,5n)\ce{^{135}Ce} reaction.
}
\label{fig:135CE}
\end{figure}

The measured cross sections are plotted in Fig. \ref{fig:135CE}.  While TALYS and EMPIRE approximately predicted the magnitude of this cross section, the energy at which it peaks is clearly miscalculated by the TALYS exciton model.  This is not surprising given the lack of low-lying level information available in \ce{^{135}Ce} in the angular momentum range that would be populated in a (p,5n) channel, with data from only a single EC-decay of \ce{^{135}Pr} and a pair of (HI,xn) measurements \cite{1972AR08, 1990MA26, 2005JAZZ}.  It is worth noting that all three models predict similar intensities for both the (p,6n) and (p,5n) channels, which together account for more than 15\% of the total reaction cross section.  Due to the strong feeding of the (p,5n) channel up to 65 MeV, a higher incident energy beam (at least 70 MeV) would be required to produce \ce{^{134}Ce} for medical applications without this major contaminant.

\subsection{\ce{^{139}La}(p,3n)\ce{^{137m,g}Ce} Cross Sections}

The decays of both the 34.4 hour isomer and the 9.0 hour ground state in \ce{^{137}Ce} were observed, which allows for the measurement of the independent cross sections (i.e. the isomer to ground state branching ratio) for this reaction.  

\begin{figure}[htb]
\includegraphics[width=9cm]{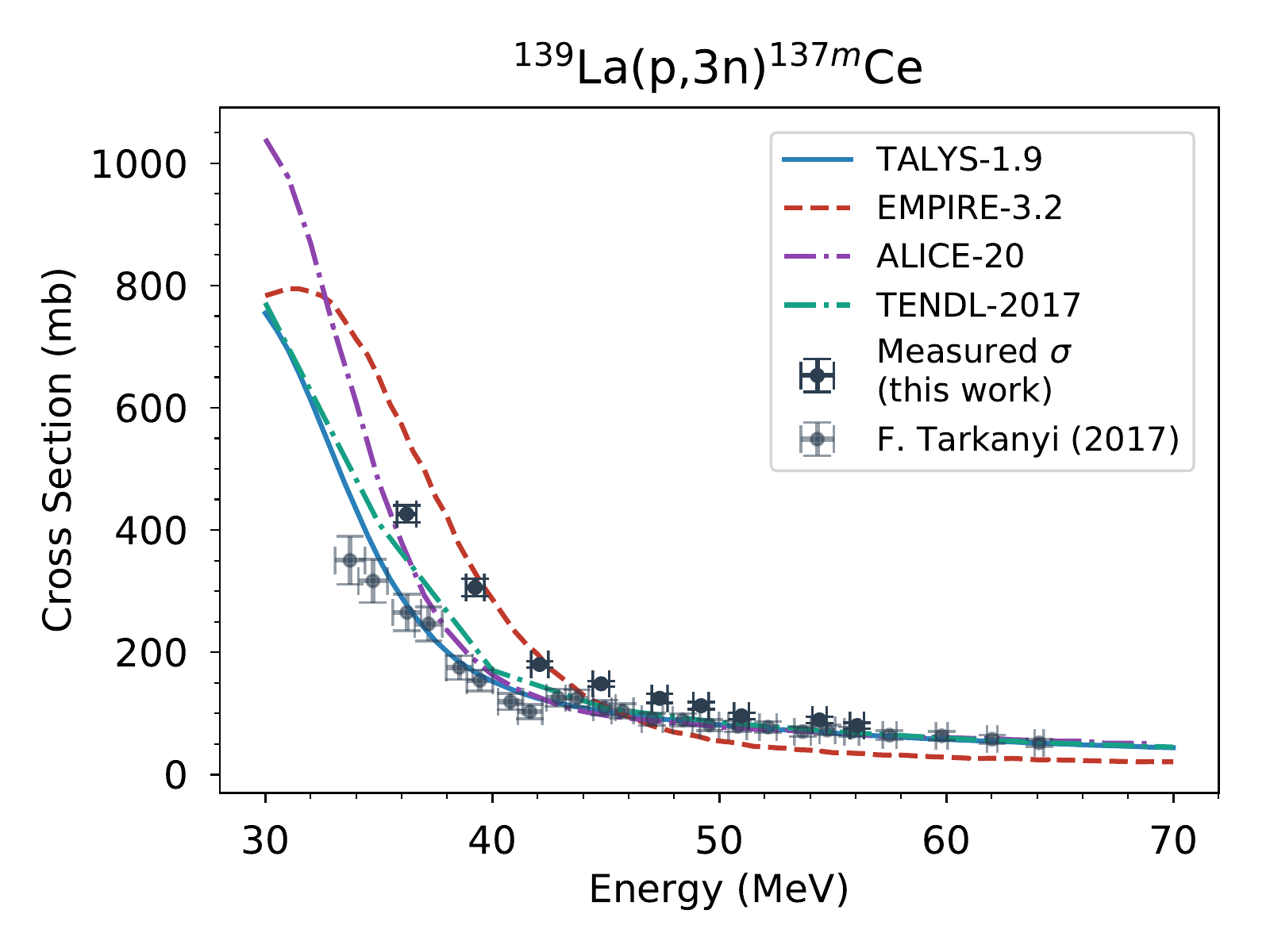}
\caption{Measured cross sections for the \ce{^{139}La}(p,3n)\ce{^{137m}Ce} reaction.
}
\label{fig:137CEm}
\end{figure}

\begin{figure}[htb]
\includegraphics[width=9cm]{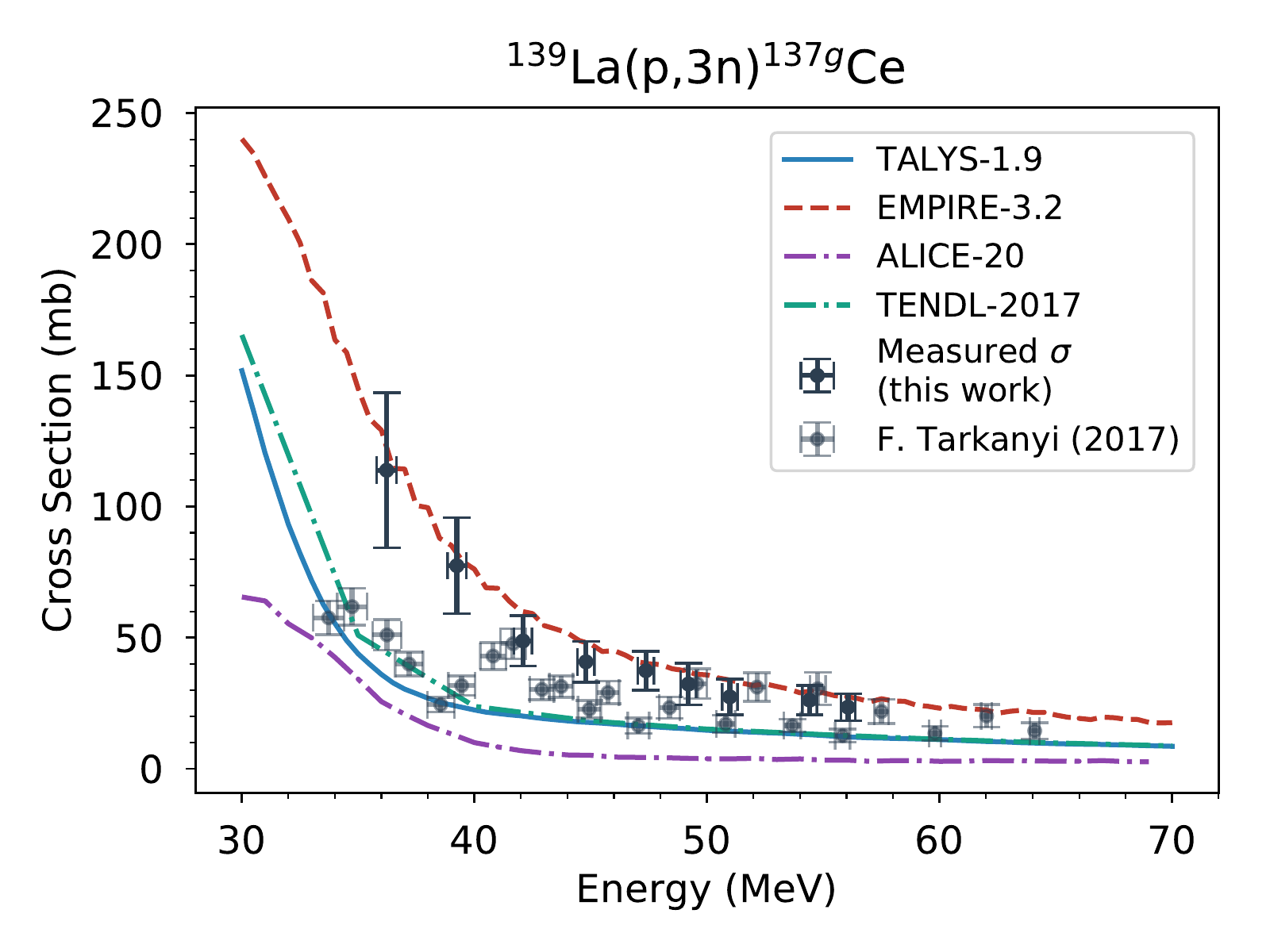}
\caption{Measured cross sections for the \ce{^{139}La}(p,3n)\ce{^{137g}Ce} reaction.
}
\label{fig:137CEg}
\end{figure}

The measured cross sections for the \ce{^{139}La}(p,3n)\ce{^{137m}Ce} reaction are plotted in Fig. \ref{fig:137CEm} and the \ce{^{139}La}(p,3n)\ce{^{137g}Ce} cross sections are plotted in Fig. \ref{fig:137CEg}.  In neither case is there a clear ``best fit'' among the models.  Production of both the isomer and ground state shows better agreement with modeling codes than in the (p,5n) and (p,6n) reactions, and possibly an adjustment of the level density model or spin-cutoff parameters would bring the calculations into agreement with the data \cite{BRINK1957215}.

\subsection{\ce{^{139}La}(p,n)\ce{^{139}Ce} Cross Section}

Measurement of the direct reaction (p,n) was possible using the 80\%, 165.85 keV $\gamma$ emission from the \ce{^{139}Ce} ground state decay.  This is reported as a cumulative cross section measurement, as the feeding from the short lived isomer \ce{^{139m}Ce} (t$_{1/2}$=58 s) could not be quantified before it had completely decayed away.

\begin{figure}[htb]
\includegraphics[width=9cm]{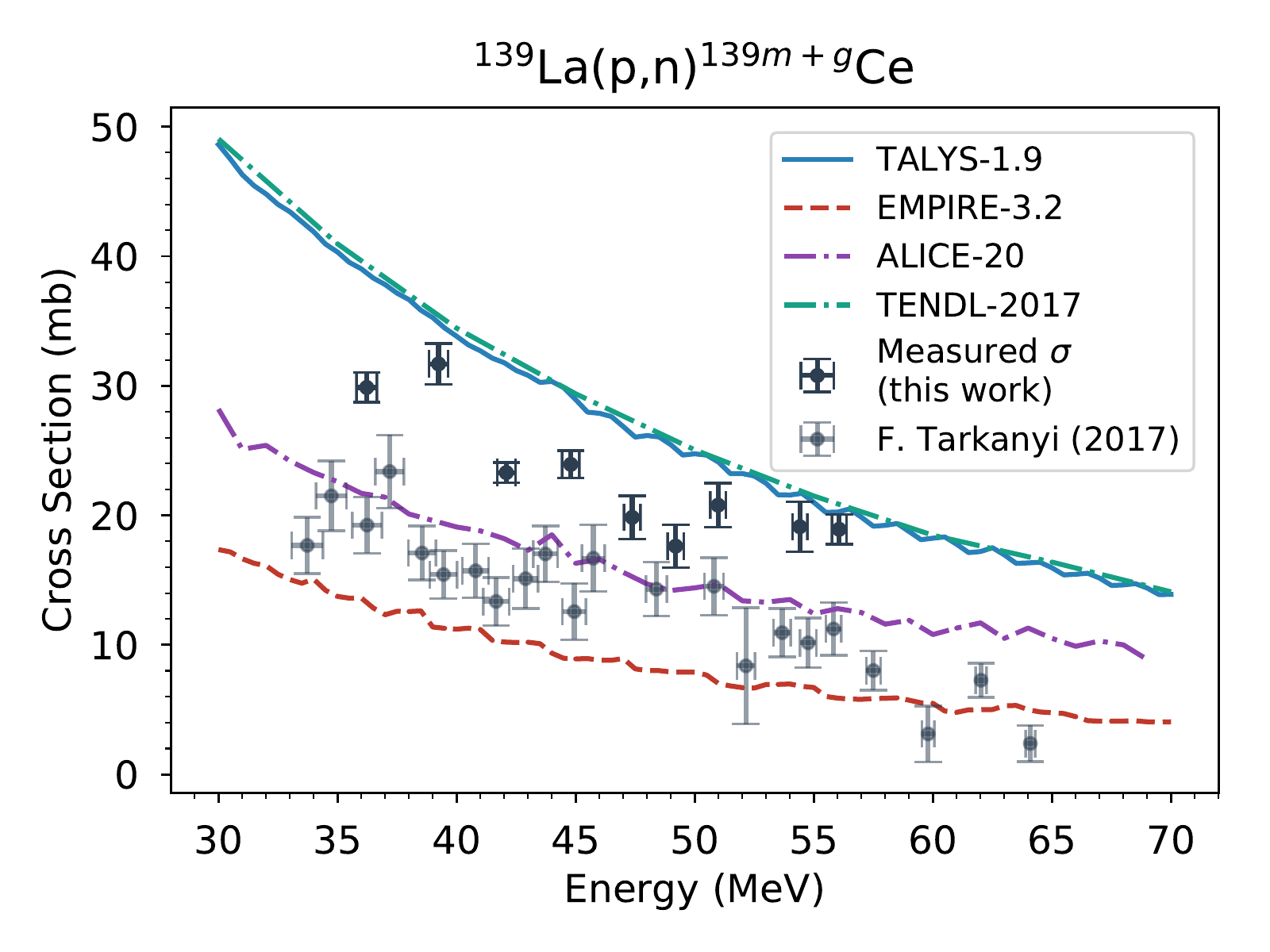}
\caption{Measured cross sections for the \ce{^{139}La}(p,n)\ce{^{139}Ce} reaction.
}
\label{fig:139CE}
\end{figure}

The measured cross sections for the \ce{^{139}La}(p,n)\ce{^{139}Ce} reaction are plotted in Fig. \ref{fig:139CE}.  All three models reproduce the shape of the excitation function, with ALICE being the most accurate at predicting the overall magnitude.

\subsection{\ce{^{139}La}(p,x)\ce{^{132}Cs} Cross Section}

Despite only producing activities on the order of 1--2 Bq, the \ce{^{139}La}(p,x)\ce{^{132}Cs} reaction was measurable because its 6.48 day activity was longer lived than of most isotopes measured in this study, and because the 97.59\%, 667.71 keV $\gamma$ line was well isolated and could be counted for multiple days.  This provides an opportunity to study the ability of the models to predict exit channels that represent a smaller component of the overall reaction cross section.   

\begin{figure}[htb]
\includegraphics[width=9cm]{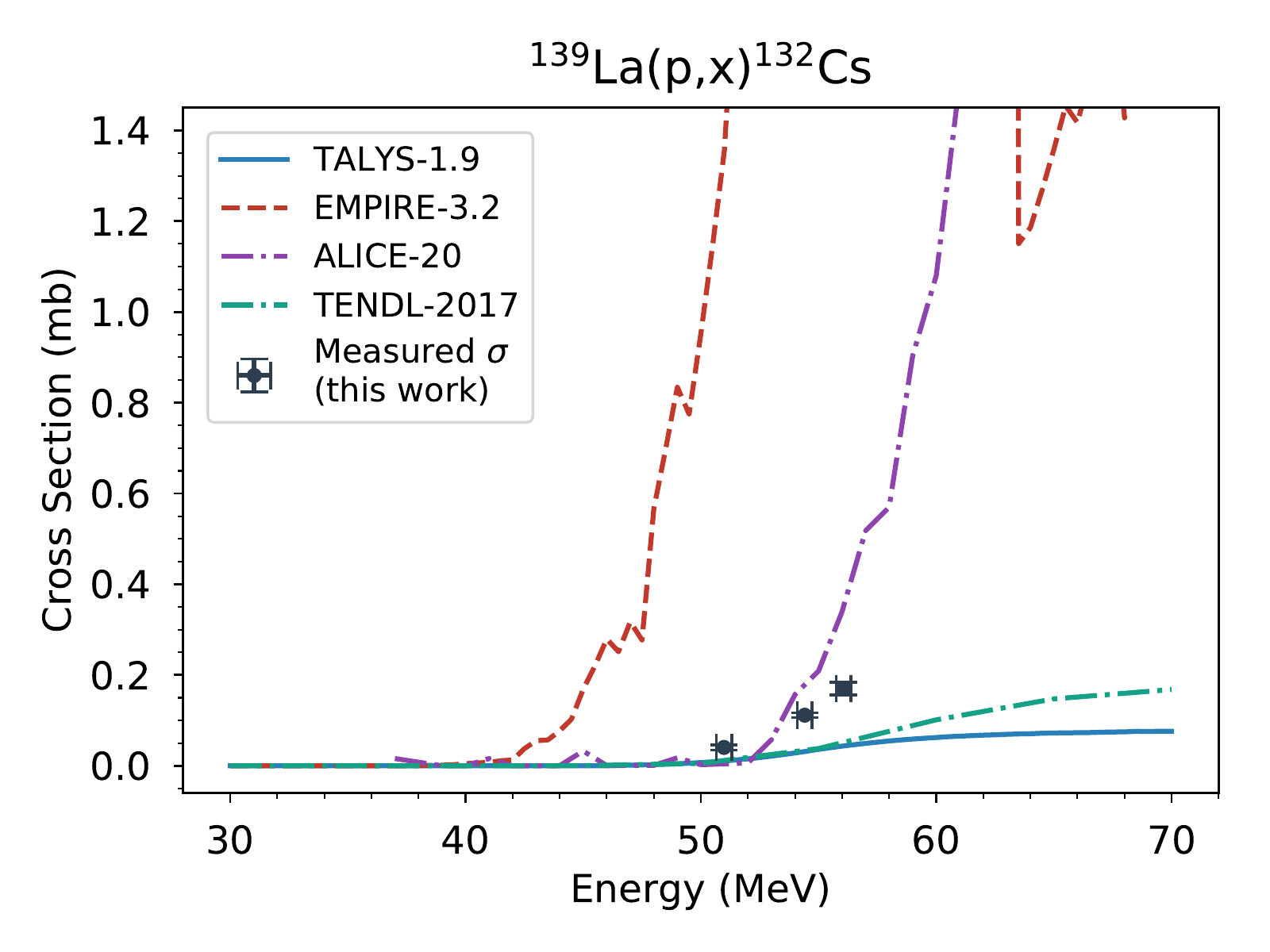}
\caption{Measured cross sections for the \ce{^{139}La}(p,x)\ce{^{132}Cs} reaction.
}
\label{fig:132CS}
\end{figure}

Fig. \ref{fig:132CS} plots the measured \ce{^{139}La}(p,x)\ce{^{132}Cs} cross sections.  EMPIRE over-predicted this cross section by almost a factor of 100, whereas the TALYS calculation was far more consistent with measurements, with the ALICE prediction in between the two.  The significant discrepancies seen in EMPIRE calculations are common in weakly-fed reaction channels like this one ($<$0.1\% of total cross section), whose behavior are extremely sensitive to more dominant channels \cite{KONING2003231}.

\subsection{\ce{^{139}La}(p,x)\ce{^{133m,g}Ba} Cross Sections}

Another reaction where the decays of both an isomer (J$_{\pi}$=11/2$^-$) and the ground state (J$_{\pi}$=1/2$^+$) were observed was the \ce{^{139}La}(p,x)\ce{^{133}Ba} exit channel.  The main challenge in this measurement was to identify the \ce{^{133}Ba} ground state decays, which had low activities ($\approx$ 0.1 Bq) due to the 10.55 year half-life of that isotope, and due to contaminating peaks in the spectrum for the first few weeks after the irradiation.  Fortunately, the isomer has a strong peak (17.69\%) at 275.92 keV, which allowed its activity to be measured with $<$1\% uncertainty.  Multiple long counts enabled the identification of the \ce{^{133}Ba} ground state and separation of the ground state activity due to the population of the isomer.  Neither \ce{^{133}Ce} (t$_{1/2}$=97 (4) m) nor \ce{^{133}La} (t$_{1/2}$=3.912 (8) h) were observed in this work, both of which emit strong characteristic $\gamma$ lines which would have been observable with the HPGe detector used here.  Therefore, the cross sections for the \ce{^{139}La}(p,x)\ce{^{133m,g}Ba} reactions are reported in this work as independent.

\begin{figure}[htb]
\includegraphics[width=9cm]{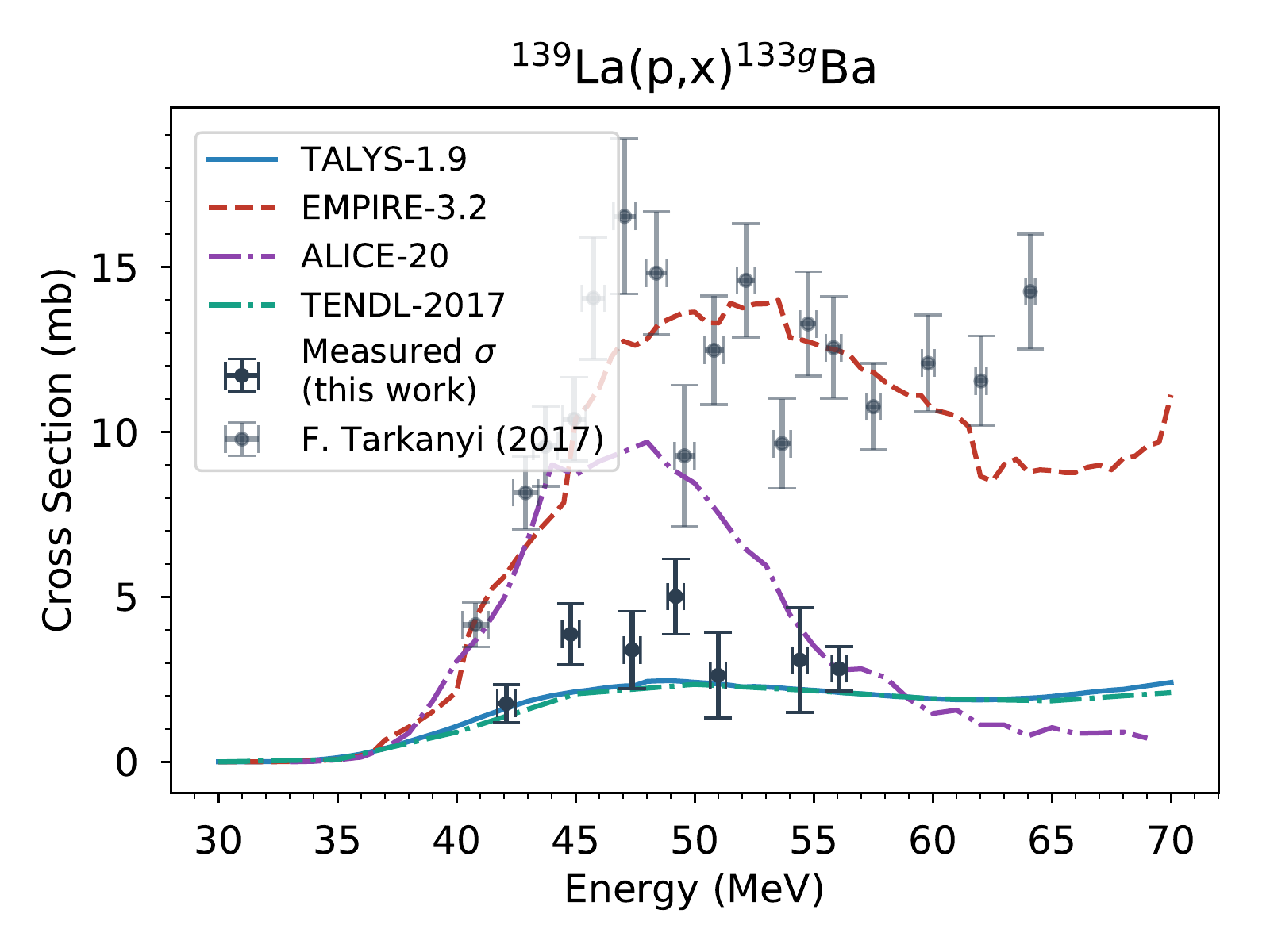}
\caption{Measured cross sections for the \ce{^{139}La}(p,x)\ce{^{133g}Ba} reaction.
}
\label{fig:133BAg}
\end{figure}

Fig. \ref{fig:133BAg} plots the measured \ce{^{139}La}(p,x)\ce{^{133g}Ba} reaction cross sections.  The relative uncertainties were very large due to the long half-life (10.55 y) \cite{A133} and the weak feeding of this channel.

\begin{figure}[htb]
\includegraphics[width=9cm]{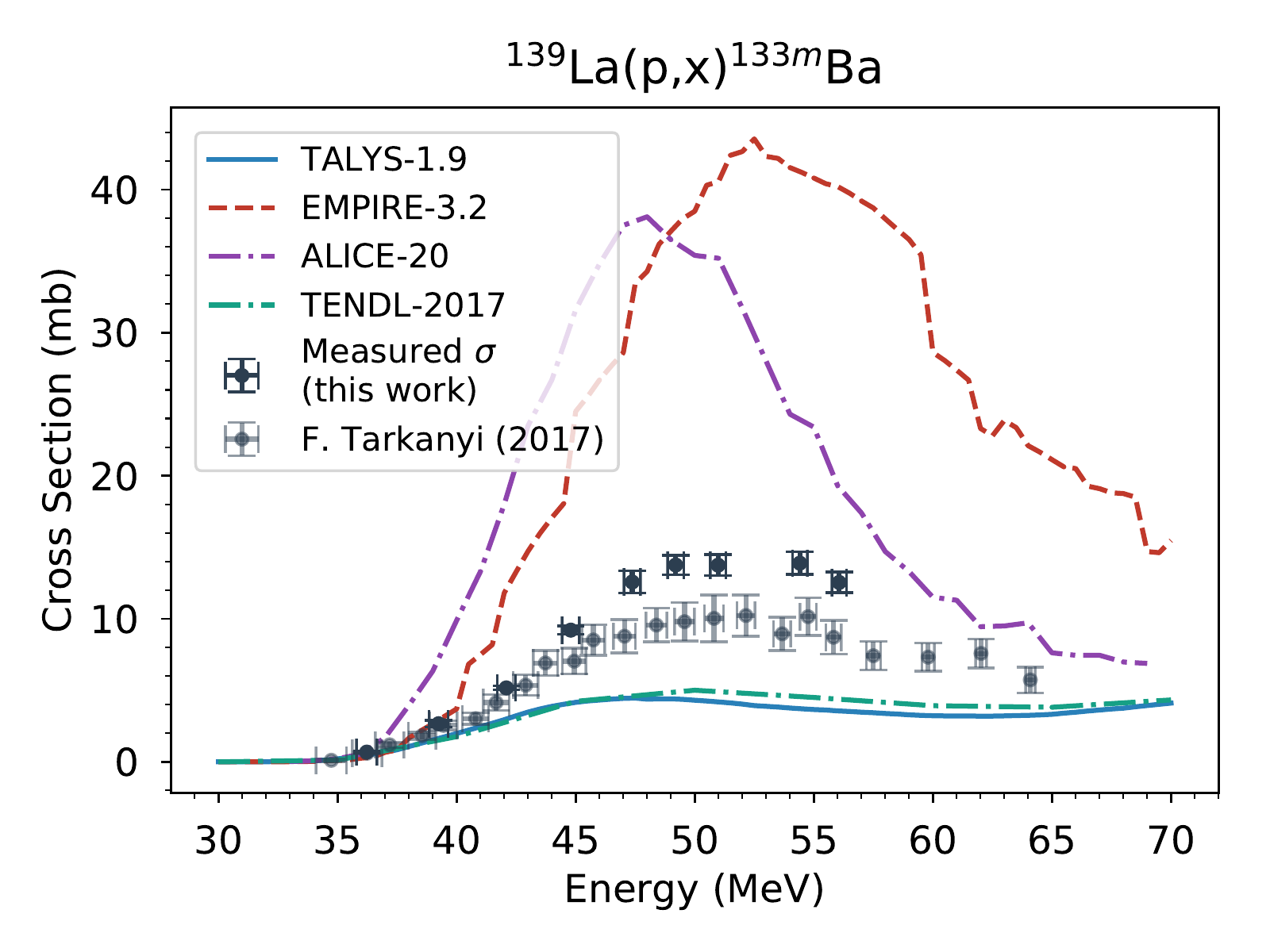}
\caption{Measured cross sections for the \ce{^{139}La}(p,x)\ce{^{133m}Ba} reaction.
}
\label{fig:133BAm}
\end{figure}

Fig. \ref{fig:133BAm} plots the measured cross sections for the \ce{^{139}La}(p,x)\ce{^{133m}Ba} reaction.  This measurement was much more precise due to the better counting statistics from the 275.925 keV line.  Here again the results from EMPIRE were in better agreement with the location of the ``compound peak'' as a function of energy, although none of the three codes accurately predicted the magnitude of this relatively modest exit channel.

\subsection{\ce{^{139}La}(p,x)\ce{^{135}La} Cross Section}

The final cross section measured in the Lanthanum stack was in the \ce{^{139}La}(p,x)\ce{^{135}La} reaction, which has relatively weak $\gamma$ emissions but was able to be identified using the 1.52\%, 480.51 keV $\gamma$ line.  The uncertainties in this measurement were $\approx$30\% because the EoB \ce{^{135}La} activities were small compared to the in-feeding from \ce{^{135}Ce} decay.  %  The uncertainties in this measurement were affected by the in-feeding from the decay of \ce{^{135}Ce}, but because the activity of \ce{^{135}Ce} was able to be measured very precisely ($\approx$1\% error), this had only a small effect on the uncertainty of the resulting cross section.

\begin{figure}[htb]
\includegraphics[width=9cm]{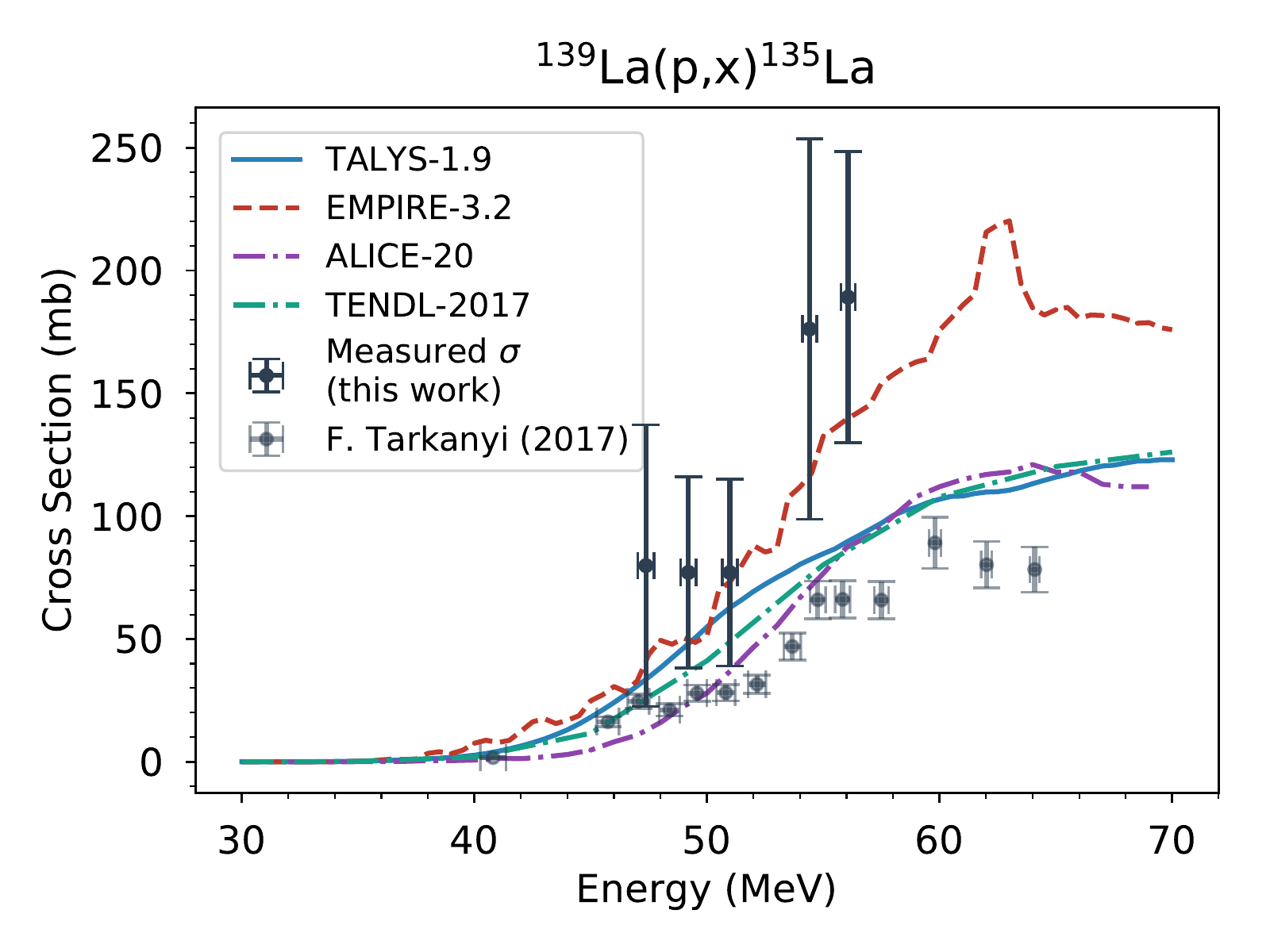}
\caption{Measured cross sections for the \ce{^{139}La}(p,x)\ce{^{135}La} reaction.
}
\label{fig:135LA}
\end{figure}

Fig. \ref{fig:135LA} plots the measured cross sections for the \ce{^{139}La}(p,x)\ce{^{135}La} reaction.  The EMPIRE model once again predicts a larger magnitude than the other codes for this channel, however there is not a clear ``best fit'' among the three codes, particularly because of the large uncertainties in the measurements of this work.
%There are some fluctuations in the EMPIRE predictions, which are likely due to the non-deterministic nature of the pre-equilibrium model it is using \cite{HERMAN20072655}.

\subsection{\ce{^{nat}Cu}(p,x)\ce{^{61}Cu} Cross Section}

\begin{figure}[htb]
\includegraphics[width=9cm]{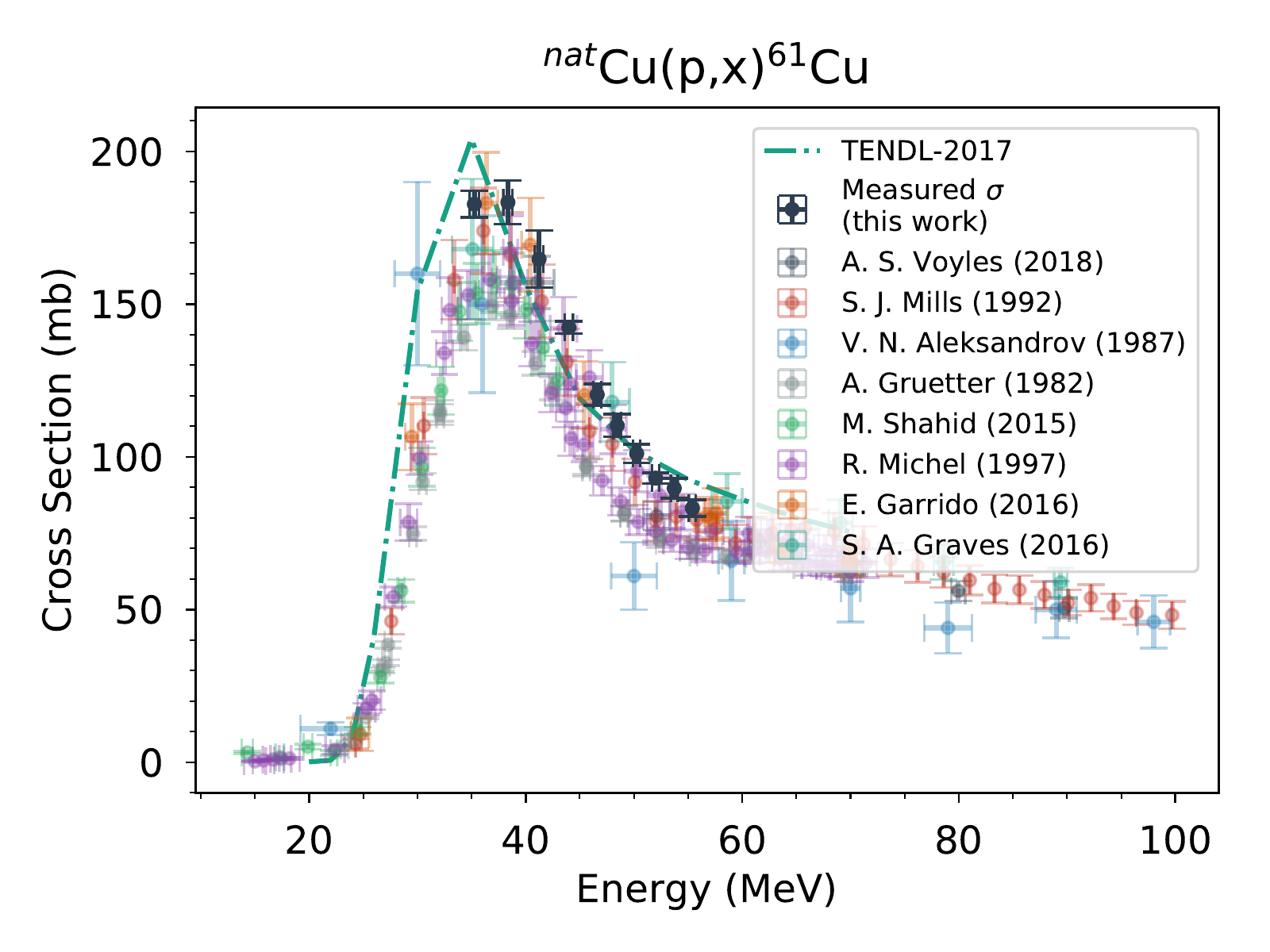}
\caption{Measured cross sections for the \ce{^{nat}Cu}(p,x)\ce{^{61}Cu} reaction \cite{Grutter, Aleksandrov, Mills, Michel, Shahid, GRAVES201644, Garrido, VOYLES201853}. 
}
\label{fig:61CU}
\end{figure}

In addition to their use in proton current determination, $\gamma$ spectroscopy of the copper monitor foils provided a measurement of the \ce{^{nat}Cu}(p,x)\ce{^{61}Cu} reaction through the observation of the 282.9 keV $\gamma$ line (12.2\%) in \ce{^{61}Cu} (t$_{1/2}$=3.339 (8) h).  These measurements are plotted in Fig. \ref{fig:61CU}, in comparison with literature data retrieved from the EXFOR database \cite{Grutter, Aleksandrov, Mills, Michel, Shahid, GRAVES201644, Garrido, VOYLES201853}.  This measurement is consistent with literature data compiled in EXFOR, both in the shape and magnitude of the excitation function, which builds confidence in the energy and current assignments determined in this work as well as the overall measurement and data reduction methodology.  Because the cross sections in this experiment are measured relative to the 2017 IAEA-recommended monitor cross sections, this measurement may be particularly useful if the \ce{^{nat}Cu}(p,x)\ce{^{61}Cu} reaction were to be included in a future evaluation, which may be unlikely due to the potential for secondary neutron contamination in this channel.

\section{Summary and Conclusions}
In this experiment, we measured the cross sections for nine \ce{^{139}La}(p,x) reactions using a 57 MeV proton beam stacked-target irradiation at the LBNL 88-Inch Cyclotron.  These measurements are compared with the outputs of the TALYS, EMPIRE and ALICE nuclear reaction modeling codes, using default parameters.  In many cases, all three codes had difficulty reproducing the magnitude of the cross sections, but TALYS consistently under-predicted the energy of the ``compound peak'', whereas the EMPIRE and ALICE predictions tended to better reproduce the shape of the excitation functions.  Better agreement with the models was found for the more strongly-fed exit channels.  This illustrates the current deficiencies in  reaction modeling of pre-equilibrium particle emission, which are highly sensitive to the nuclear level density and spin-distribution models employed. This systematic issue will be the subject of a forthcoming publication.

Particular emphasis was placed on the production of \ce{^{134}Ce}, a radionuclide with applications as a positron-emitting analogue of \ce{^{225}Ac}, a promising medical radionulcide.  The results of this study show that in order to produce significant quantities of \ce{^{134}Ce} from the \ce{^{139}La}(p,6n) reaction, a proton beam of higher energy would be more effective.  The highest-energy proton beam available at the LBNL 88-Inch Cyclotron (60 MeV) produces unacceptable quantities of other long-lived cerium  radionuclides, which must be avoided for biodistribution studies.  Based on the present work, we believe that a proton beam of at least 70 MeV will be required to produce significant activities of \ce{^{134}Ce} without major contaminants.

\section*{Acknowledgements}
We wish to acknowledge our thanks to the operators of the 88-Inch Cyclotron, Brien Ninemire, Nick Brickner, Tom Gimpel and Scott Small, for their efforts in setting a new ``high-water mark'' for the maximum proton energy extracted from the machine as well as for their assistance and support.  We would also like to thank the members of the LBNL Nuclear Data group and the Nuclear Engineering department at UC Berkeley, who contributed their time and knowledge towards the review of this experiment.

This work has been performed under the auspices of the U.S. Department of Energy by Lawrence Berkeley National Laboratory under contract No. LAB16-1588 NSD.  This research is supported by the U.S. Department of Energy Isotope Program, managed by the Office of Science for Nuclear Physics.

\bibliography{writeup}

\appendix
\section{Stack Design}
\label{stack_appendix}
\ \ 
\begin{ruledtabular}
\begin{tabular}{cccc}
Foil Id & Compound & $\Delta x$ (mm) & $\rho \Delta x$ (mg/cm$^2$) \Bstrut \\
\hline
\Tstrut SS3 & 316 SS & 0.13 & 100.48 $\pm$ 0.46 \\
La01 & La & 0.0275 & 14.59 $\pm$ 0.69 \\
Al01 & Al & 0.027 & 6.58 $\pm$ 0.02 \\
Cu01 & Cu & 0.029 & 22.13 $\pm$ 0.07 \\
E1 & Al & 0.254 & 68.53 $\pm$ 5.08 \\
La02 & La & 0.0278 & 15.55 $\pm$ 0.71 \\
Al02 & Al & 0.0278 & 6.67 $\pm$ 0.12 \\
Cu02 & Cu & 0.0293 & 22.23 $\pm$ 0.44 \\
E2 & Al & 0.254 & 68.53 $\pm$ 5.08 \\
La03 & La & 0.0315 & 15.12 $\pm$ 0.83 \\
Al03 & Al & 0.027 & 6.7 $\pm$ 0.03 \\
Cu03 & Cu & 0.031 & 22.24 $\pm$ 0.07 \\
E3 & Al & 0.254 & 68.53 $\pm$ 5.08 \\
La04 & La & 0.0288 & 14.95 $\pm$ 0.66 \\
Al04 & Al & 0.027 & 6.68 $\pm$ 0.03 \\
Cu04 & Cu & 0.0317 & 22.49 $\pm$ 0.42 \\
E4 & Al & 0.254 & 68.53 $\pm$ 5.08 \\
La05 & La & 0.027 & 15.07 $\pm$ 0.65 \\
Al05 & Al & 0.027 & 6.64 $\pm$ 0.01 \\
Cu05 & Cu & 0.0313 & 22.39 $\pm$ 0.42 \\
E5 & Al & 0.254 & 68.53 $\pm$ 5.08 \\
La06 & La & 0.026 & 14.32 $\pm$ 0.78 \\
Al06 & Al & 0.0278 & 6.66 $\pm$ 0.23 \\
Cu06 & Cu & 0.031 & 22.22 $\pm$ 0.05 \\
E6+E7 & Al & 0.508 & 137.06 $\pm$ 10.16 \\
La07 & La & 0.0258 & 14.21 $\pm$ 0.29 \\
Al07 & Al & 0.0273 & 6.64 $\pm$ 0.12 \\
Cu07 & Cu & 0.031 & 22.4 $\pm$ 0.05 \\
E8+E9 & Al & 0.508 & 137.06 $\pm$ 10.16 \\
La08 & La & 0.0283 & 15.64 $\pm$ 0.28 \\
Al08 & Al & 0.0273 & 6.72 $\pm$ 0.13 \\
Cu08 & Cu & 0.032 & 22.16 $\pm$ 1.2 \\
E10+E11 & Al & 0.508 & 137.06 $\pm$ 10.16 \\
La09 & La & 0.0268 & 12.67 $\pm$ 0.51 \\
Al09 & Al & 0.0275 & 6.65 $\pm$ 0.14 \\
Cu09 & Cu & 0.031 & 22.2 $\pm$ 0.72 \\
E12+E13 & Al & 0.508 & 137.06 $\pm$ 10.16 \\
La10 & La & 0.0278 & 16.14 $\pm$ 0.3 \\
Al10 & Al & 0.027 & 6.73 $\pm$ 0.02 \\
Cu10 & Cu & 0.031 & 22.5 $\pm$ 0.05 \\
SS4 & 316 SS & 0.13 & 101.26 $\pm$ 0.79 \\\end{tabular}
\label{table:stack}
\end{ruledtabular}
\ \

\section{Relevant Nuclear Data \cite{A134, A135, A137, A133, A139, A132, A61, A62, A63, A58, A22, A24, unpub137}}
\label{nudat_appendix}
\ \ 
\begin{ruledtabular}
\begin{tabular}{cccc}
Isotope & $\gamma$ Energy (keV) & $I_{\gamma}$ (\%) & $T_{1/2}$ \Bstrut \\
\hline
\Tstrut\ce{^{134}Ce} & - & - & 3.16 (4) d \\
\ce{^{134}La} & 604.721 (2) & 5.04 (20) & 6.45 (16) m \\
\ce{^{135}Ce} & 265.56 (2) & 41.8 (14) & 17.7 (3) h \\
\ce{^{137m}Ce} & 254.29 (5) & 11.1 (4) & 34.4 (3) h \\
\ce{^{137g}Ce} & 447.15 (8) & 1.22 (3) & 9.0 (3) h \\
\ce{^{139g}Ce} & 165.8575 (11) & 79.95 (6) & 137.64 (2) d \\
\ce{^{135}La} & 480.51 (2) & 1.52 (24) & 19.5 (2) h \\
\ce{^{133m}Ba} & 275.925 (7) & 17.69 (25) & 38.93 (1) h \\
\ce{^{133g}Ba} & 356.0129 (7) & 62.05 (19) & 10.551 (11) y \\
\ce{^{132}Cs} & 667.714 (2) & 97.59 (9) & 6.480 (6) d \\
\ce{^{61}Cu} & 282.956 (10) & 12.2 (22) & 3.339 (8) h \\
\ce{^{62}Zn} & 596.56 (13) & 26.0 (20) & 9.193 (15) h \\
\ce{^{63}Zn} & 669.62 (5) & 8.2 (3) & 38.47 (5) m \\
\ce{^{58}Co} & 810.7593 (20) & 99.45 (1) & 70.86 (6) d \\
\ce{^{22}Na} & 1274.537 (7) & 99.940 (14) & 2.6018 (22) y \\
\ce{^{24}Na} & 1368.626 (5) & 99.9936 (15) & 14.997 (12) h \\
\end{tabular}
\label{table:decay_data}
\end{ruledtabular}
\ \

\end{document}